\documentclass[10pt,journal,compsoc,x11names]{IEEEtran}
\usepackage[switch]{lineno}
\usepackage{amsfonts}
\usepackage{amsmath, bm}
\usepackage{xspace}
\usepackage{enumitem}
\usepackage{amssymb}
\usepackage{booktabs}
\usepackage[hyphens]{url}

\usepackage{booktabs}  % for \toprule \midrule \bottomrule
\usepackage{tabularx}  % for tabularx and X columns
\usepackage{array}     % for \newcolumntype and \arraybackslash
\newcolumntype{Y}{>{\raggedright\arraybackslash}X} % ragged-right X

\usepackage{ragged2e}
\usepackage{hyperref}
\usepackage{multirow}
\usepackage{graphicx}
\usepackage{xcolor}
\usepackage{color}
\usepackage{colortbl}
\usepackage{tablefootnote}
\usepackage{pifont}
\usepackage{makecell}
\usepackage[most]{tcolorbox}
\usepackage{framed}
\usepackage{mdframed}
\usepackage{subfigure}
\usepackage{caption}
\usepackage{longtable}
\usepackage{float}
\usepackage{booktabs}
\usepackage{graphicx}
\usepackage{tikz}
\usepackage[edges]{forest}
\definecolor{hidden-draw}{RGB}{0,51,102}
\definecolor{hidden-blue}{RGB}{194,232,247}

\tikzstyle{my-box}=[
    rectangle,
    draw=hidden-draw,
    rounded corners,
    text opacity=1,
    minimum height=1.5em,
    minimum width=5em,
    inner sep=2pt,
    align=center,
    fill opacity=.5,
]
\tikzstyle{leaf}=[my-box, minimum height=1.5em,
    fill=hidden-blue!60, text=black, align=left,font=\scriptsize,
    inner xsep=2pt,
    inner ysep=4pt,
]

\newcolumntype{H}{>{\setbox0=\hbox\bgroup}c<{\egroup}@{}}

\newcommand{\ignore}[1]{}

\definecolor{gold}{RGB}{205,133,63}
\definecolor{fGreen}{RGB}{34,139,34}
\definecolor{tOrange}{RGB}{255,215,0}
\definecolor{tBlue}{RGB}{135,206,250}
\definecolor{tPink}{RGB}{255,204,204}
\definecolor{tGreen}{RGB}{205,230,199}
\definecolor{tGold}{RGB}{255,215,0}

\usepackage{cite}
\usepackage[numbers,sort&compress]{natbib}

\ifCLASSINFOpdf
\else
\fi

\begin{document}
\title{\fontsize{21pt}{30pt}\selectfont A Survey on Code Generation with LLM-based Agents}

\author{Yihong Dong*, Xue Jiang*, Jiaru Qian*, Tian Wang, Kechi Zhang, Zhi Jin, and Ge Li
\IEEEcompsocitemizethanks{
\IEEEcompsocthanksitem Version: v2 (September 30, 2025).
\IEEEcompsocthanksitem GitHub link: \mbox{\url{https://github.com/JiaruQian/awesome-llm-based-agent4code}} 
\IEEEcompsocthanksitem * The authors contribute equally to this work. 
\IEEEcompsocthanksitem 
The authors are mainly with the School of Computer Science, Peking University, Beijing, China; \\
Contact e-mail: {dongyh, jiangxue}@stu.pku.edu.cn; 
}
}

\markboth{}%
{Shell \MakeLowercase{\textit{et al.}}: Bare Advanced Demo of IEEEtran.cls for IEEE Computer Society Journals}

\IEEEtitleabstractindextext{%
\begin{abstract}
\justifying
Code generation agents powered by large language models (LLMs) are revolutionizing the software development paradigm. Distinct from previous code generation techniques, code generation agents are characterized by three core features. 1) Autonomy: the ability to independently manage the entire workflow, from task decomposition to coding and debugging. 2) Expanded task scope: capabilities that extend beyond generating code snippets to encompass the full software development lifecycle (SDLC). 3) Enhancement of engineering practicality: a shift in research emphasis from algorithmic innovation toward practical engineering challenges, such as system reliability, process management, and tool integration. This domain has recently witnessed rapid development and an explosion in research, demonstrating significant application potential. This paper presents a systematic survey of the field of LLM-based code generation agents. We trace the technology's developmental trajectory from its inception and systematically categorize its core techniques, including both single-agent and multi-agent architectures. Furthermore, this survey details the applications of LLM-based agents across the full SDLC, summarizes mainstream evaluation benchmarks and metrics, and catalogs representative tools. Finally, by analyzing the primary challenges, we identify and propose several foundational, long-term research directions for the future work of the field.
\end{abstract}

\begin{IEEEkeywords}
Code Generation; Software Development; Large Language Models; LLM-based Agent; Multi-agent System
\end{IEEEkeywords}}

\maketitle

\IEEEdisplaynontitleabstractindextext

\IEEEpeerreviewmaketitle

\section{Introduction}

Code generation aims to automatically transform human intentions expressed in certain specifications into executable computer programs, serving as a fundamental approach to improving software productivity. Early research in code generation primarily adopted program synthesis methods~\cite{kitzelmann2009inductive}, deriving verifiably correct programs through formal specifications. However, due to the difficulty of writing specifications, this approach was long confined to well-defined specific tasks. To enhance generalization capabilities, research subsequently shifted toward data-driven paradigms based on deep learning, treating code generation as a probabilistic sequence learning problem~\cite{ling2016latent,yin2017syntactic}. Nevertheless, code snippets generated by such methods often had limited functionality and frequently contained syntactic or semantic errors, leading to compilation or execution failures. Consequently, automated program writing has long been considered an extremely challenging task. This situation fundamentally changed with the emergence of Large Language Models (LLMs)~\cite{touvron2023llama,touvron2023llama2,ouyang2022training}. Although LLM technology originated from natural language processing, it has also demonstrated remarkable potential in code generation tasks. This is primarily attributed to the massive code contributions from open-source communities represented by GitHub in the vast pre-training corpora, enabling models to master the syntax and semantics of programming languages as well as programming algorithms and paradigms. Benefiting from this, the field of code generation has ushered in unprecedented development opportunities and rapidly become one of the most promising application directions for LLM technology.

Although LLM-based code generation techniques have shown excellent performance in generating standalone programs~\cite{roziere2023code,wang2021codet5,li2022competition}, their single-response mode exposes significant limitations when handling complex, engineering-oriented software development tasks. Native LLMs lack the ability to autonomously decompose tasks, interact with real development environments, validate generated code, and implement continuous self-correction mechanisms. As a result, they struggle to independently complete software development tasks that require cross-file context understanding, dynamic debugging, and iterative optimization. To address these problems, LLM-based code generation agents have been proposed, which use LLMs as the brain to construct code agents capable of autonomous planning, action, observation, and iterative optimization~\cite{jiang2024self,le2023codechain,pan2025codecor}. They are capable of simulating the complete workflow of human programmers, including analyzing requirements, writing code, running tests, diagnosing errors, and applying fixes~\cite{rasheed2024codepori,manish2024autonomous}. This comprehensive capability enables them to handle complex programming tasks that exceed the limitations of individual LLMs, thereby producing higher-quality and more reliable software outputs. Such advancements represent a critical step toward achieving higher levels of automation in software development.
This pursuit is closely aligned with the historical evolution of software engineering.
Over time, development paradigms have shifted from individual programming to team-based collaborative development.
This transition has been driven by the growing complexity of software systems and the corresponding need for a clear division of labor.

LLM-based code generation has been extensively discussed and surveyed, while systematic research on code generation agents remains insufficient. To stimulate exploration in this field, we emphasize three core distinctions between code generation agents and previous code generation techniques. First is autonomy. Traditional code generation models assist human developers through code completion or function generation in a passive manner. In contrast, code generation agents can actively manage and execute development workflows from requirements to implementation~\cite{dong2024self,qian2023chatdev,hong2023metagpt}. This paradigm transforms the role of the developer from code writer to task definer, process supervisor, and final result reviewer. The second distinction is the expansion of code generation task scope. Previous code generation research typically involved tasks with clear boundaries and well-defined specifications, such as completing code lines based on context and generating function bodies based on function signatures~\cite{lu2021codexglue}.    Code generation agents, however, can cover most tasks in software development, including handling ambiguous requirements~\cite{mu2023clarifygpt}, implementing entire project coding~\cite{dong2024self,qian2023chatdev,hong2023metagpt}, testing and refactoring programs~\cite{wang2024xuat,zhang2025logiagent,baumgartner2024ai}, and iterative optimization based on real-time feedback~\cite{le2023codechain}. Finally, there is a shift in research focus from algorithmic innovation to engineering practice. Early code generation research focused primarily on improving algorithmic accuracy, such as enhancing model architectures or optimizing training methods to pursue higher code syntactic correctness and semantic matching~\cite{yin2017syntactic,rabinovich2017abstract,guo2020graphcodebert,guo2022unixcoder}. The research focus of code generation agents has significantly shifted toward engineering implementation, including how to ensure agent reliability~\cite{hu2025qualityflow,pan2025codecor}, how to manage complex workflows~\cite{hong2023metagpt,ishibashi2024self}, and how to enable agents to efficiently invoke external tools~\cite{zhang2024codeagent,wang2024toolgen}. These problems have expanded from pure model generation capabilities to system design, process management, and human-computer collaboration, entering territories closer to classical software engineering.

Today, LLM-based code generation agents are profoundly impacting software development. Tools represented by Claude Code and Cursor can already preliminarily complete end-to-end software development through multi-agent collaborative division of labor, typically with lower time and cost than human teams. Vibe Coding has become a popular programming practice, where developers use natural language prompts to describe problems and provide them to specialized LLMs for software development, which then generate software~\cite{sapkota2025vibe}. However, some important problems remain unresolved. First, integrating code generation agents with real development environments faces numerous difficulties. Actual software projects often contain large and private codebases, customized build processes, internal API calling specifications, and unwritten team conventions. How to enable agents to efficiently understand and utilize this non-public, highly contextualized information is a critical challenge that must be solved for their transition from general demonstrations to professional tools. Then, code generated by agents often contains logical defects, performance pitfalls, or security vulnerabilities that are difficult to cover with unit tests~\cite{schafer2023adaptive,xu2025multi,dearing2025leveraging,peng2025sysllmatic,shi2024harnessing,foster2025mutation}, forcing developers to invest unexpected effort in code review and manual repairs.

Facing opportunities and challenges, we need more attention to the research and development of code generation agents. To this end, this survey comprehensively reviews relevant literature, introduces concepts and knowledge of LLM-based code generation agents, and systematically reviews the latest advances in code generation agents, mainly covering key technologies, evaluation methods, deployed tools, and various applications throughout the software development lifecycle, aiming to provide a solid knowledge foundation for researchers in this field. We understand that there have been previous English survey articles on LLM-based agents for software engineering~\cite{jin2024llms,liu2024large,he2025llm,wang2024agents}, which focus on classifying technologies from an application perspective. However, we believe that despite varying application scenarios, the underlying technical methods are common. Therefore, this work differs from existing surveys in two ways: First, we classify and introduce from a methodological perspective, providing relatively in-depth technical references for research and development of code generation agents; Second, given the rapid development of this field, we focus on integrating the latest research advances and challenges faced.

The structure of this paper is organized as follows: We first introduce relevant background knowledge, technological development, and core concepts. Then, we deeply explore key technologies of code generation agents (covering both single-agent and multi-agent systems); subsequently, we discuss specific applications and evaluation methods of these code generation agents in software development and catalog representative tools in the market; finally, the article analyzes current challenges in the field and prospects for future directions, providing a summary of the entire work.

\section{Literature Collection and Technological Development Trend Analysis}

\subsection{Literature Collection}

To ensure the completeness and comprehensiveness of the research, this study employed systematic literature retrieval methods to collect all relevant high-quality literature as comprehensively as possible. This research selected authoritative academic databases, including ACM Digital Library, IEEE Xplore, SpringerLink, Google Scholar, DBLP Computer Science Bibliography, and China National Knowledge Infrastructure (CNKI), to conduct literature retrieval, ensuring the comprehensiveness and authority of the retrieval scope. The literature retrieval time span covered from 2022 to June 2025, encompassing important development stages and the latest advances in this research field.

We adopted a bilingual retrieval strategy in both Chinese and English. The Chinese and English search keyword combination was "('Code Generation' OR 'Software Development') AND ('LLM' OR 'Large Language Model' OR 'Large Model' OR 'Language Model') AND ('Agent' OR 'Multi-agent' OR 'Agentic')". The retrieval fields covered titles, abstracts, keywords, and index terms. To ensure literature quality, the retrieval scope was limited to top-tier international academic conferences and journals in software engineering and artificial intelligence recommended by the China Computer Federation (CCF), including ICSE, ISSTA, ASE, FSE, TOSEM, TSE, ACL, ICML, ICLR, AAAI, and other international top-tier conferences and journals, as well as CCF-A Chinese journals such as Journal of Computer Research and Development, Journal of Software, Science China Information Sciences, and Chinese Journal of Computers.

Additionally, given the rapid development of this field, to track the latest research developments, this study also collected high-quality works published on preprint platforms such as arXiv. This research further expanded the retrieval scope through forward and backward snowball searching for each piece of literature. Through the above retrieval strategies, 447 candidate literature pieces were initially obtained.

We screened the retrieved literature with five screening criteria: (1) Exclude duplicately published papers and multiple versions of similar content from the same research team to avoid content redundancy. (2) Manually evaluate literature from preprint platforms like arXiv, retaining only works with significant academic impact\footnote{The citation count of the literature exceeds the citation threshold for hot papers in the SCI computer science discipline during the same period \url{https://esi.clarivate.com/ThresholdsAction.action}.} and innovation. (3) Exclude books, dissertations, and conference short papers, retaining only complete academic papers published in authoritative journals and conferences. (4) Focus on technical innovation papers, excluding pure technical reports, empirical surveys, and review literature. (5) Conduct in-depth content analysis and manual review of initially screened literature, eliminating literature with low relevance to the research topic. The above screening rules ensured high quality and high relevance of the literature collection: Rule (1) avoided research content redundancy; Rules (2)(3)(4) focused on core technological breakthroughs and frontier developments in the field; Rule (5) ensured high matching between literature and research objectives. After strict screening, 100 high-quality core literature pieces were finally obtained, constituting the main analysis objects of this research.

\subsection{Technological Development Trend Analysis}

\begin{figure*}[t]
    \centering
    \subfigure[Distribution of papers published by year\label{fig:paper_distribution_a}]{
        \includegraphics[width=0.45\textwidth]{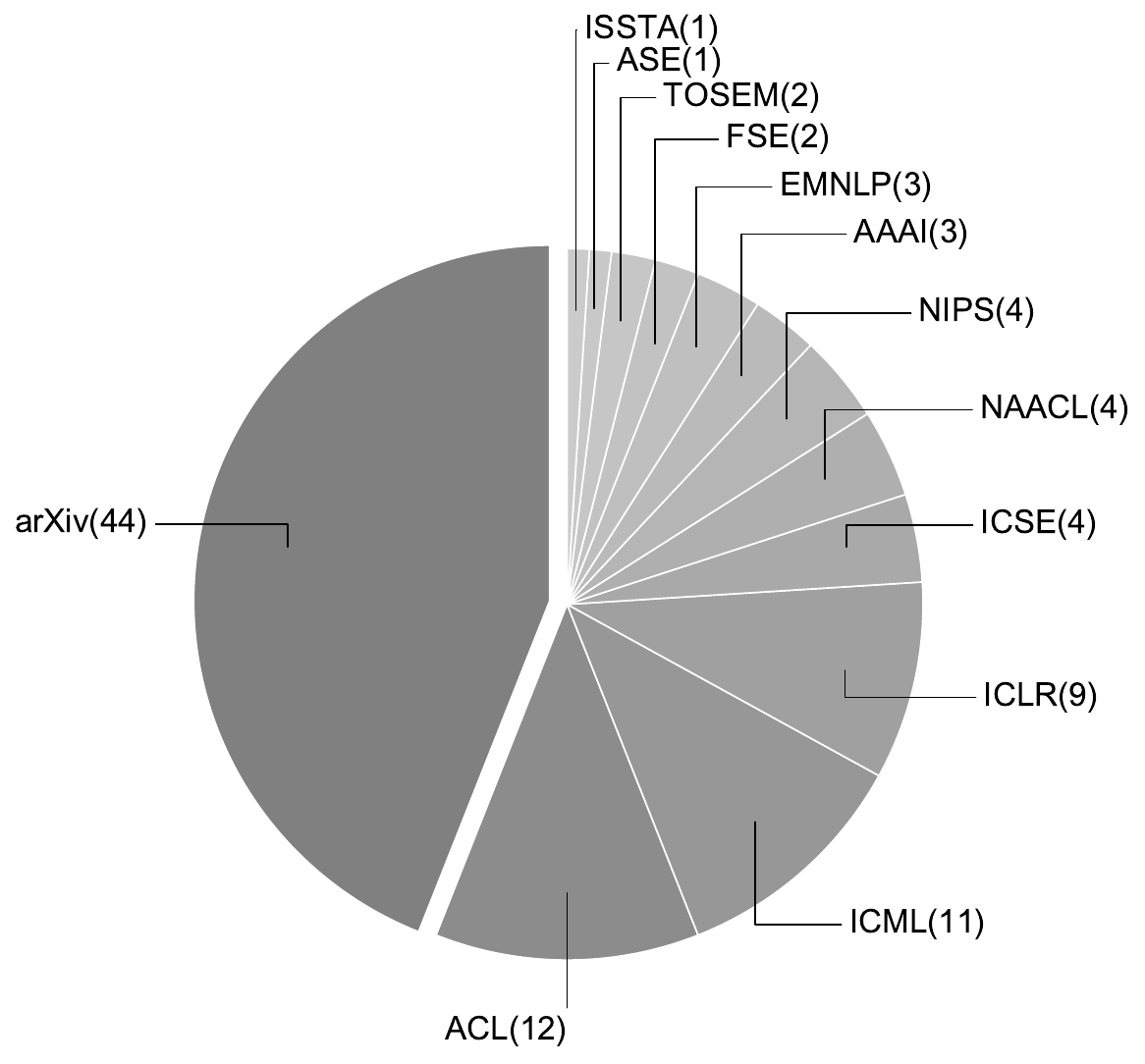}
    }
    \hfill
    \subfigure[Distribution of papers published by conference or journal\label{fig:paper_distribution_b}]{
        \includegraphics[width=0.45\textwidth]{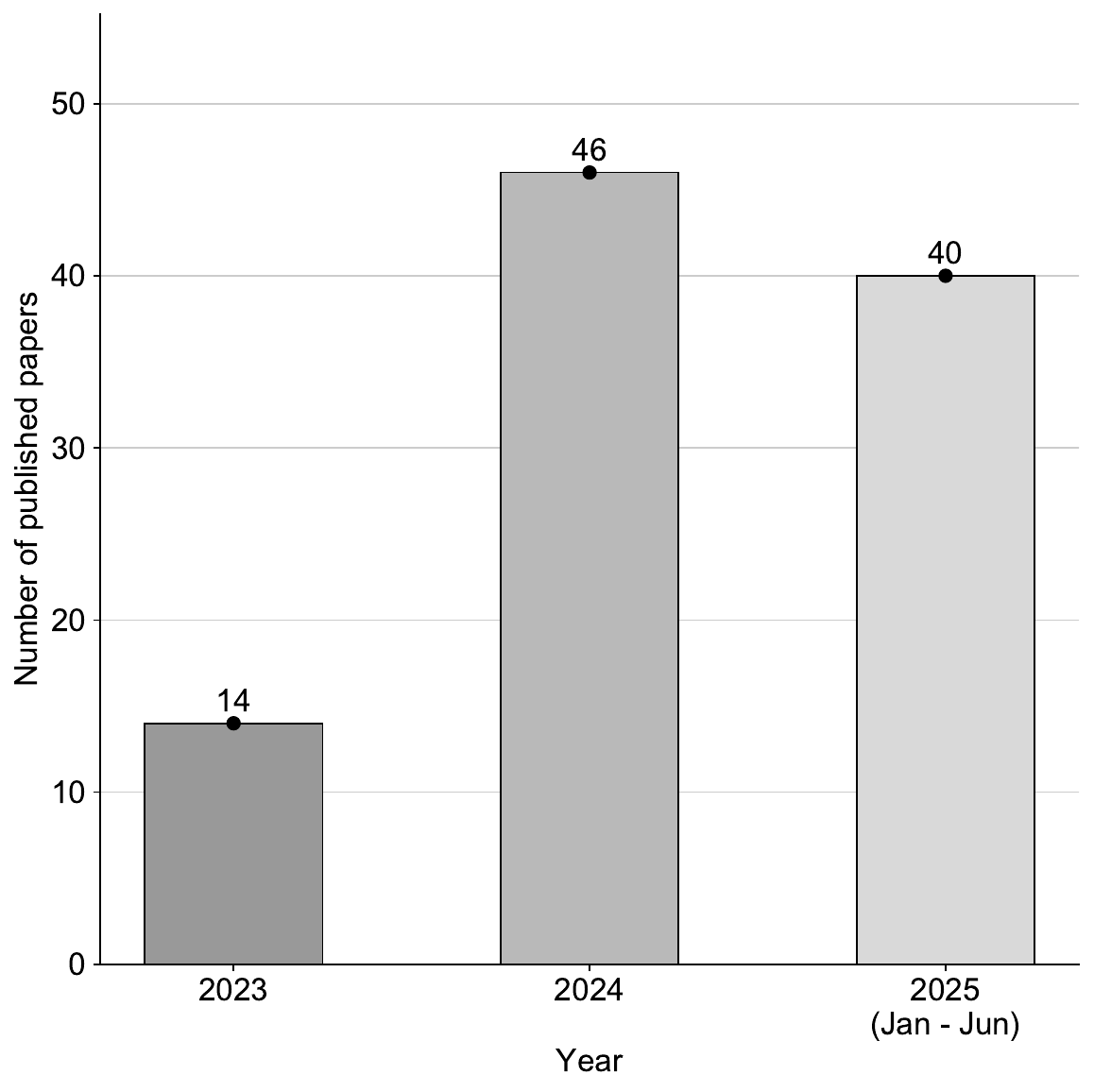}
    }
    \caption{Distribution of papers related to LLM-based code generation agents.}
    \label{fig:paper_distribution}
\end{figure*}

Figure~\ref{fig:paper_distribution} shows the distribution of papers related to LLM-based code generation agents. Figure~\ref{fig:paper_distribution_a} shows statistics on the number of related papers published in different years. From the data in Figure~\ref{fig:paper_distribution_a}, it can be found that the number of papers published in this field shows a year-over-year increasing trend. Since its emergence in 2023, code generation agents have shown tremendous potential, and subsequently, the attention and research enthusiasm from academia and industry have rapidly increased, with the number of related studies also growing significantly. Figure~\ref{fig:paper_distribution_b} lists statistics on the number of related papers published in top-tier academic conferences or journals each year since the emergence of code generation agent technology. Papers related to code generation agents appear not only in top-tier software engineering conferences and journals (such as ICSE, ASE, FSE, ISSTA, TOSEM) but are also frequently accepted by mainstream conferences in natural language processing and artificial intelligence (such as ACL, ICLR, NeurIPS, ICML, AAAI), indicating that related research has received widespread attention from multiple disciplinary fields. Additionally, due to the rapid technological development in this field and certain lag in traditional paper review cycles, many research results are published as preprints on arXiv, many of which have received high citation counts and have had important impacts on field development.

\section{Background Knowledge}

\subsection{Code Generation in Software Engineering}

Code generation~\cite{li2022competition,svyatkovskiy2020intellicode,herrington2003code} refers to the process of automatically converting structured or unstructured input information (i.e. natural language requirement descriptions, system design documents, existing code snippets, etc.) into source code. Its essence is to map abstract intentions or task objectives to specific programming implementations, aiming to reduce manual coding costs, improve development efficiency, and to some extent reduce human errors, ultimately achieving the goal of software development automation. This task typically requires that the generated results are syntactically legal, semantically consistent with expectations, and capable of running correctly in the target environment.

The complexity and task diversity of modern software development pose new requirements for code generation technology to handle open and complex needs. Meanwhile, the industry urgently expects to comprehensively integrate code generation methods into development processes to improve efficiency. Against this background, traditional code generation techniques encounter three fundamental limitations~\cite{becker2023programming,xu2022ide,huynh2025large}.
First, they lack sufficient contextual understanding, rendering them ineffective in processing open-ended or high-level instructions.
Second, their generative capacity is limited, making it challenging to produce logically coherent and functionally complete code.
Third, they exhibit poor generality and flexibility, hindering their ability to adapt to diverse software development tasks.
In recent years, the emergence of large language models has opened new avenues for addressing these challenges.

\subsection{Large Language Models}

Large Language Models (LLMs)~\cite{kasneci2023chatgpt,chang2024survey,zhao2023surveyllm,naveed2023comprehensive} are a class of pre-trained language models based on deep learning technology, primarily using the Transformer model~\cite{vaswani2017attention} as the core architecture. The basic idea is to learn statistical patterns, semantic structures, and contextual relationships in language through unsupervised pre-training on massive text corpora. The core task of LLMs is typically conditional language modeling, i.e., predicting subsequent tokens given context, formally expressed as maximizing conditional probability $P(x_t|x_1,x_2,\cdots,x_{t-1})$. This mechanism enables LLMs to possess powerful language understanding and generation capabilities.

In the field of code generation, since training data contains a large amount of high-quality open-source code libraries and programming documentation, LLMs can learn and master syntactic rules of multiple programming languages as well as common programming paradigms, while understanding the mapping relationship between natural language descriptions and code logic, thereby achieving generation of executable code from natural language descriptions, or intelligent completion and refactoring in existing contexts. Typical code generation LLMs include Codex~\cite{chen2021evaluating}, CodeLlama~\cite{rozière2024codellamaopenfoundation}, DeepSeek-Coder~\cite{guo2024deepseek}, and Qwen2.5-Coder~\cite{hui2024qwen2}, etc., which have been widely applied in software engineering scenarios such as code completion, test code generation, and bug fixing, demonstrating powerful code generation and understanding capabilities.

In addition to basic language modeling capabilities, LLMs also exhibit various important emergent abilities. First is planning capability, where LLMs can generate natural language plans to guide subsequent generation~\cite{wei2022chain,wang2024planning}. Second is tool usage capability, where LLMs can actively invoke external tools based on task requirements to enhance problem-solving capabilities. By encapsulating available tool API calls, existing work has been able to integrate various external tools, such as search engines~\cite{nakano2022webgptbrowserassistedquestionansweringhuman}, calculators~\cite{schick2023toolformer}, and compilers~\cite{gao2023palprogramaidedlanguagemodels}, etc., to further expand the capability boundaries of LLMs. Third is environmental interaction capability, where LLMs have the ability to receive feedback from external environments and execute operations according to instructions, enabling perception, decision-making, and action in dynamic environments \cite{park2023generative,mehta2023improving} . 

\begin{figure*}[t!]
  \centering
  \includegraphics[width=0.96\textwidth]{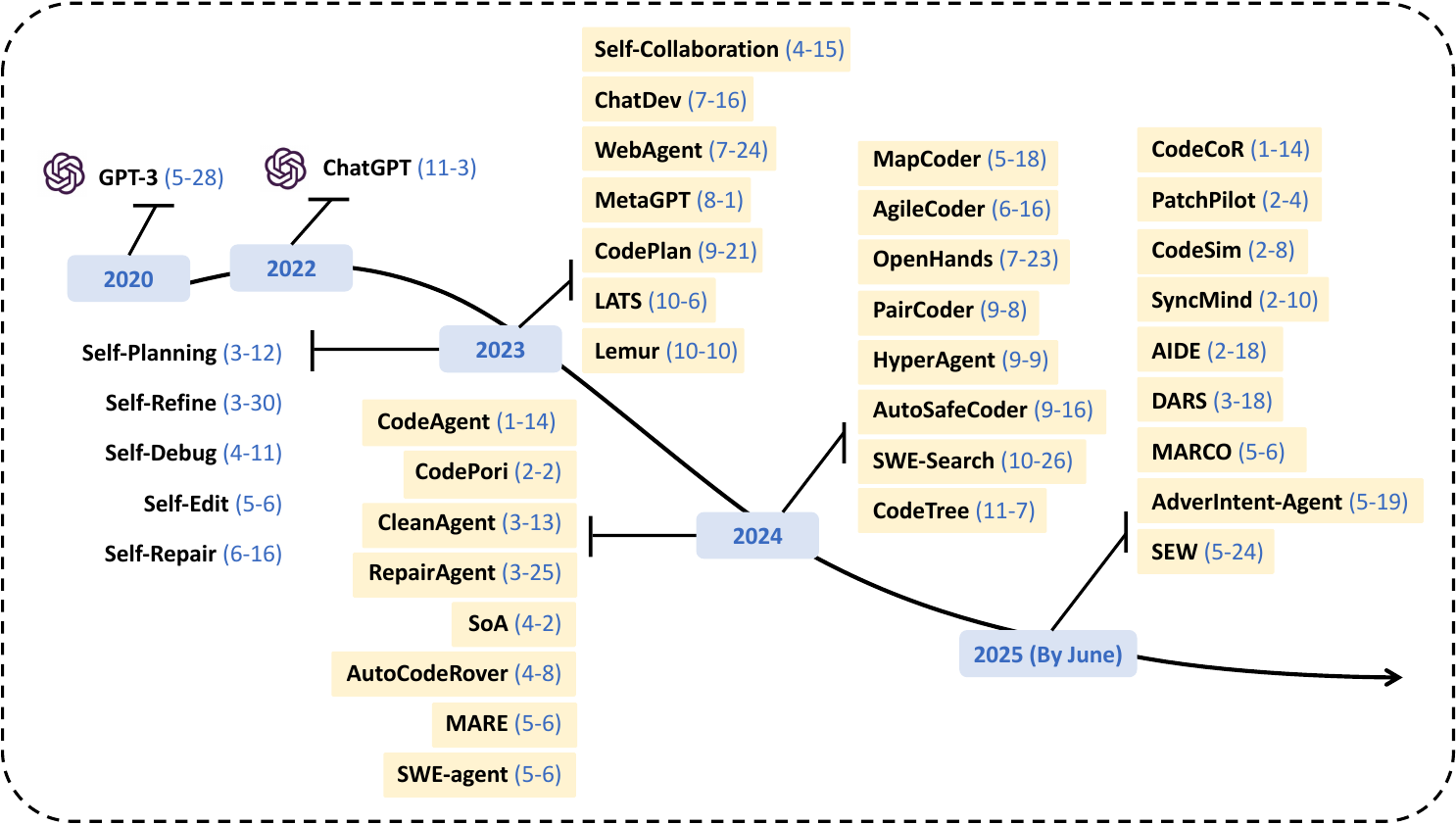}
  \caption{Evolution of key technologies in LLM-based code generation agents, where works not marked in yellow are important prerequisite technologies.}
  \label{fig:work_timeline}
\end{figure*}

\subsection{LLM-based Agents}

LLM-based Agents~\cite{xi2025rise,wang2024surveyagents,guo2024large,cheng2024exploring} are a class of system architectures that use LLMs as the core reasoning engine, integrating perception, memory, decision-making, and action modules to enable models with autonomous task execution capabilities. Agents, as a fundamental paradigm in artificial intelligence, initially focused primarily on perception-decision-execution loops under reinforcement learning frameworks, emphasizing the interaction process between models and environments and long-term reward optimization. With the widespread success of LLMs in natural language reasoning tasks, researchers began attempting to use them as decision cores to construct LLM-based agents. The language modeling capability of LLMs enables them to directly process unstructured text instructions, understand complex semantic intentions, and in the absence of explicit supervisory signals, autonomously organize and execute tasks by combining environmental perception, language planning, and tool invocation. Due to the complexity, modularity, and collaborative nature of software engineering, the code generation field was among the first to see applications of LLM-based agents.

LLM-based agents mainly include core components such as planning, memory, tool usage, and reflection. The \textbf{planning component} is responsible for task decomposition, breaking large tasks into smaller, manageable sub-goals, thereby efficiently handling complex tasks. The \textbf{memory component} is divided into short-term memory and long-term memory, where short-term memory is implemented through the context window of LLMs. Through prompt engineering~\cite{giray2023prompt,white2023prompt}, information directly related to the current task is placed into this working memory, thereby guiding the model's immediate reasoning and behavior. Long-term memory breaks through the capacity limitations of context windows by constructing external persistent knowledge bases. Current mainstream technical implementations adopt the Retrieval Augmented Generation (RAG)~\cite{gao2023retrieval,lewis2020retrieval} framework, encoding massive information into high-dimensional vectors and storing them in dedicated vector databases. When agents need to invoke historical experience or domain knowledge, they can quickly locate and extract relevant information through efficient vector similarity retrieval algorithms. The \textbf{reflection component} allows agents to examine, evaluate, and correct their own generated content or existing data, thereby improving past action decisions and correcting previous errors for continuous improvement. The \textbf{tool usage component} interacts with external physical or digital environments, enabling agents to transcend their native model limitations. Language models themselves are closed systems with knowledge cutoff dates and lack the ability to perform computations or invoke external APIs. The tool usage module greatly enhances their action space by granting agents permission to invoke external functions or APIs.

From the perspective of composition and interaction complexity, agent systems can be divided into single-agent and multi-agent categories.

\begin{itemize}
    \item \textbf{Single Agent}: Refers to an independent centralized agent that autonomously completes all tasks through its inherent planning, tool usage, and reflection capabilities. There is no complexity of inter-agent interaction.

    \item \textbf{Multi-Agent}: Refers to systems composed of multiple heterogeneous or homogeneous agents. In multi-agent systems, goals are achieved through communication, collaboration, and negotiation between agents. In multi-agent systems, role-based professional division of labor is a common collaborative enhancement strategy. By assigning specific roles to different agents (such as "analyst", "programmer", "tester"), problems that far exceed the capabilities of individual agents in scale and complexity can be solved.
\end{itemize}

\subsection{Differences Between LLMs and LLM-based Agents}

LLMs and LLM-based agents exhibit architectural and capability differences when solving code generation tasks~\cite{jin2024llms}, specifically:

The core advantage of LLMs lies in their powerful contextual generation capabilities~\cite{li2023largeincontext,wei2023larger}, enabling them to efficiently predict and output semantically coherent code snippets based on given inputs. However, their operation is essentially a single, passive response process. Models receive inputs, generate outputs based on their pre-trained knowledge, and the entire process lacks active planning, state maintenance, or continuous interaction with external environments. This limits their performance when handling complex, ambiguous, or multi-step collaborative software development tasks.

LLM-based agents, on the other hand, construct a dynamic workflow with autonomy, interactivity, and iterativity, integrating the generation capabilities of LLMs into an intelligent system capable of active planning, execution, observation, and adjustment. In the field of code generation, LLM-based agents not only utilize the generation capabilities of LLMs but also endow them with capabilities for task decomposition~\cite{huang2023agentcoder,phan2024hyperagent}, tool invocation (such as compilers, API documentation queries)~\cite{zhang2023toolcoder,phan2024repohyper}, and self-correction based on feedback (such as execution errors, user feedback)~\cite{madaan2023self,zhang2023self}. LLMs serve as reasoning engines in agent frameworks, responsible for making decisions based on current environmental states and determining next actions, thereby driving tasks to gradual completion.

\section{Key Technologies and Methods}

This chapter introduces the key technologies and methods of LLM-based code generation agents, divided into two main categories: single-agent code generation methods and multi-agent code generation systems. Single-agent methods form the foundation for building multi-agent systems. We will first introduce three key technologies in single-agent systems, including planning and reasoning techniques, tool integration and retrieval enhancement, and reflection and self-improvement mechanisms. In the multi-agent code generation systems section, we will focus on collaboration mechanisms between agents, expanding from three aspects: how to arrange workflows in code generation agent systems, how to achieve efficient information interaction and management between agents, and how to transform multiple agents into a more capable overall system through collaborative optimization. We have organized the key technologies and methods chronologically, as shown in Figure~\ref{fig:work_timeline}.

\begin{figure*}[t!]
  \centering
  \includegraphics[width=0.99\textwidth]{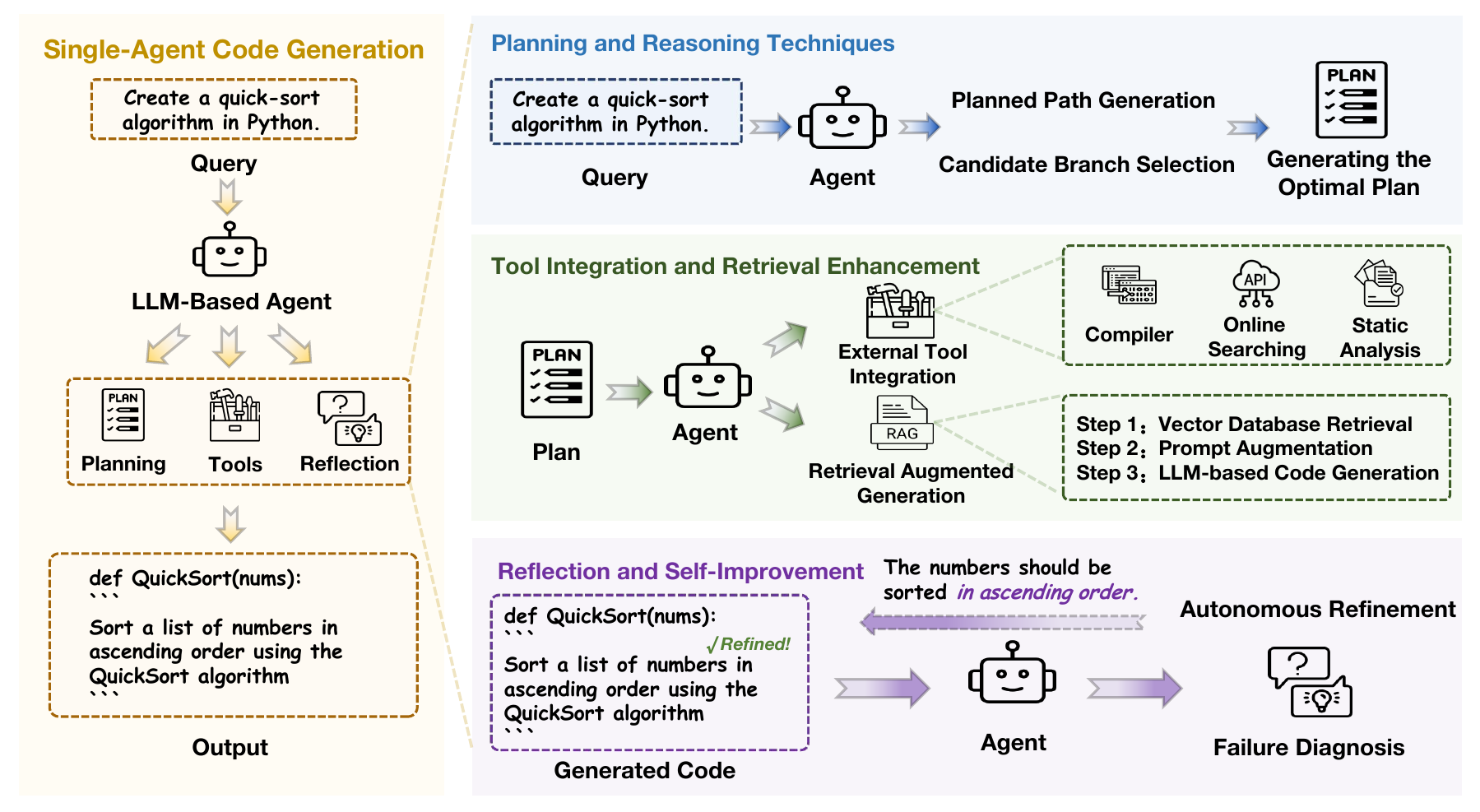}
  \caption{Overview of single-agent code generation methods}
  \label{fig:single_agent_overview}
\end{figure*}

\subsection{Single-Agent Code Generation Methods}

\subsubsection{Planning and Reasoning Techniques}

Explicit planning is widely recognized as one of the key pathways to enhance the structured reasoning capabilities of LLMs. Self-Planning~\cite{jiang2024self} was the first approach to systematically introduce the planning phase into code generation tasks.
In this framework, the model is required to first produce a sequence of high-level solution steps prior to generating any executable code.
It then implements each step according to the predefined plan, enabling the decomposition of complex problems and mitigating the overall difficulty of code synthesis.

Building on this foundation, CodeChain~\cite{le2023codechain} introduced clustering and self-revision in the planning phase, guiding the model to construct reusable modular code through multiple iterations. Subsequently, CodeAct~\cite{wang2024executable} introduces a unified action space for the agent by representing all actions as executable Python code.
By integrating a Python interpreter into the agent architecture, CodeAct enables immediate code execution, real-time feedback from the environment, and dynamic adjustment of actions through multi-turn interactions. KareCoder~\cite{huang2024knowledge} achieves external knowledge injection by integrating knowledge from libraries into the planning and reasoning process of LLMs. WebAgent~\cite{gur2023real} extended the planning mechanism to web automation scenarios, introducing a structured three-stage strategy: instruction decomposition, HTML content summarization, and program synthesis. It leverages LLMs to decompose natural language instructions into standardized sub-instructions, extract task-relevant information from lengthy HTML documents, and generate executable Python programs accordingly. Meanwhile, CodePlan~\cite{bairi2024codeplan} introduced multi-stage control flow and combined with custom control instructions, enabling the model to dynamically select "generate" or "modify" operations during the reasoning process.

Enhancing the breadth and diversity exploration capabilities of candidate solutions is another important direction in single-agent framework development. To this end, GIF-MCTS~\cite{dainese2024generating} introduced Monte Carlo Tree Search (MCTS) mechanisms into the code generation process, systematically exploring multiple potential generation paths. This method constructs decision trees through multiple sampling in each generation round, combined with execution feedback to score and filter each candidate branch, not only expanding the solution space but also significantly improving the robustness and generalization capabilities of the model in multi-solution tasks. As a representative reasoning enhancement strategy, GIF-MCTS achieved a transition from single-path planning to multi-path parallel reasoning.

Building on the above methods, PlanSearch~\cite{wang2024planning} first formalized the planning process as an explicit search task, generating multiple sets of candidate plans and conducting parallel evaluation during the reasoning phase to find optimal solutions within a larger solution space. In terms of planning structure, both CodeTree~\cite{li2024codetree} and Tree-of-Code~\cite{ni2024tree} extended previous linear structures to tree structures. CodeTree organizes the code generation process into a unified tree structure that explicitly models the sequential stages of strategy exploration, solution generation, and iterative refinement. By incorporating execution-based feedback at each stage, it enables dynamic scoring and heuristic-guided pruning to efficiently navigate the large search space. Tree-of-Code explores multiple potential paths through breadth-first generation strategies and combines execution signals for branch pruning. Subsequently, Multi-stage guided code generation for Large Language Models \cite{multi-stage} emphasize dividing the planning process into multi-stage control, introducing hierarchical objectives and intermediate reward signals to alleviate goal drift problems in end-to-end generation processes. Based on this, DARS~\cite{aggarwal2025dars} further adopts adaptive tree structures to enhance the decision process, branching new planning paths at key decision nodes of original planning paths based on code execution feedback, and dynamically selecting better planning paths by combining historical trajectories with execution results. Additionally, for hardware tasks, VerilogCoder~\cite{ho2025verilogcoder} introduced graph-structured planning mechanisms and waveform tracing tools based on abstract syntax trees, supporting structural modeling and semantic verification of Verilog code, demonstrating the adaptive potential of planning paradigms in cross-modal and domain-specific tasks. To address the challenge that traditional search methods are difficult to apply in non-serializable environments, Guided Search~\cite{zainullina2025guidedsearchstrategiesnonserializable} proposes two complementary strategies, namely one-step lookahead and trajectory selection. Both strategies are guided by a learned action-value estimator, where the model predicts potential next-step outcomes to evaluate candidate actions, and selects promising solution trajectories based on their empirical success rates. This enables effective exploration without relying on environment state serialization.

In summary, these methods reflect a shift from single-path to multi-path exploration, and from linear to structured planning. They highlight the growing importance of planning techniques in enhancing the effectiveness and flexibility of single-agent code generation approaches.

\subsubsection{Tool Integration and Retrieval Enhancement}

Integrating external tools with LLMs is one of the keys for single agents to break through their own generation capability boundaries. Building on Toolformer~\cite{schick2023toolformer}, ToolCoder~\cite{zhang2023toolcoder} proposed a code generation method that combines API search tools with LLMs. To achieve accurate tool invocation, ToolCoder automatically annotates training data, enabling the model to learn to use search tools to actively query APIs, effectively alleviating API invocation errors caused by model hallucinations. ToolGen~\cite{wang2024toolgen} integrated automatic completion tools into the code generation process of LLMs, solving dependency problems in code generation (i.e., undefined variables and member errors) through offline model fine-tuning and online automatic completion tool integration. To further enhance the ability of LLMs to handle complex requirements and complex dependencies in projects, CodeAgent~\cite{zhang2024codeagent} integrated five programming tools for the model (including website search, document reading, code symbol navigation, format checker, code interpreter), supporting interaction with software components to achieve information retrieval, code implementation, and code testing functions. In terms of tool feedback mechanisms, ROCODE~\cite{jiang2024rocode} introduces a closed-loop mechanism that integrates code generation, real-time error detection, and adaptive backtracking. During the generation process, ROCODE continuously monitors the compilation output and automatically initiates backtracking when syntax errors are detected. It further employs static program analysis to identify the minimal necessary modification scope, enabling efficient and targeted rewriting.  To enhance step-by-step control of tool invocation, CodeTool~\cite{lu2025codetool} introduced process-level supervision mechanisms, explicitly modeling and supervising each step of tool invocation processes, improving the accuracy and robustness of tool invocation, and efficiently integrating tool feedback into the generation process through incremental debugging strategies.

In recent years, Retrieval-Augmented Generation (RAG) methods have also emerged as a form of external tool invocation. RAG methods retrieve relevant information from knowledge bases or code repositories before generation to construct richer contexts, thereby alleviating limitations of knowledge, model hallucinations, and data security issues. In this direction, RepoHyper~\cite{phan2024repohyper} established repository-level vector retrieval systems, supporting the location of reusable code segments from large-scale code bases and using them as context for joint generation, effectively improving the control from the model over long-distance dependencies. To enable agents to no longer rely on pre-registered function APIs and other tools, CodeNav~\cite{gupta2024codenav} automatically indexes past real repositories based on requirements during code generation, importing relevant functions and code blocks, and adjusting based on execution result feedback. Meanwhile, AUTOPATCH~\cite{acharya2025optimizing} applied RAG to runtime performance optimization problems, combining historical code examples with control flow graph (CFG) analysis for context-aware learning, and optimizing code through contextual prompts to models. Building on this, to improve the structured expression capability of retrieval contexts, Knowledge Graph Based Repository-Level Code Generation~\cite{athale2025knowledge} proposed representing code repositories as knowledge graphs, improving retrieval quality from structural and relational perspectives, achieving over 10\% improvement in project-level code generation tasks, demonstrating the importance of structure-aware retrieval for high contextual accuracy. Furthermore, cAST~\cite{zhang2025cast} addressed the "chunking" problem in RAG pipelines by proposing a structured chunking mechanism based on abstract syntax trees (AST), improving the syntactic completeness of code retrieval through recursive partitioning and semantic coherent block merging, significantly improving metrics such as Recall and Pass@1.

In specific domains, tool integration design exhibits stronger structural constraints and domain coupling. AnalogCoder~\cite{lai2025analogcoder}, facing analog circuit design tasks, proposed encapsulating simulator functions and language models as "circuit library" invocation interfaces, and designed feedback enhancement and subcircuit reuse mechanisms, enabling the model to complete structurally reasonable circuit generation without additional training. Similarly, VerilogCoder~\cite{ho2025verilogcoder} focused on hardware code generation, achieving cross-stage logic verification and fine-grained control by integrating syntax tree-level waveform tracing tools, maintaining consistency and adjustability of generation in complex hardware logic relationships. These vertical domain integration solutions demonstrate the enormous potential of tool invocation in professional task modeling and provide transferable insights for tool fusion design in general scenarios.

In summary, single agents effectively expand their perception range and execution capabilities through external tool integration and retrieval enhancement, improving the accuracy, efficiency, and consistency of model generation, constructing tool invocation solutions from general tasks to domain-specific tasks.

\subsubsection{Reflection and Self-Improvement}

Reflection and self-improvement mechanisms offer an effective approach for enhancing code generation in single-agent systems. Unlike one-shot generation methods, these approaches enable the model to review its intermediate outputs, provide internal feedback, and iteratively refine the code during the generation process. By mimicking the human process of generating, evaluating, and revising code, they help improve the overall correctness and quality of the final output.

To address the limitations of one-step generation, such as the frequent occurrence of bugs and logical errors, Self-Refine~\cite{madaan2023self} introduces an iterative refinement framework. After generating an initial output, the model performs natural language self-evaluation to identify potential issues. Based on this feedback, it revises the output to improve quality. This approach requires no additional training or supervision and has demonstrated strong generality and effectiveness across a range of tasks, including code generation and logical reasoning. Self-Iteration~\cite{chang2023self} builds upon prior reflection-based methods by addressing the problem of error accumulation in complex code generation tasks. It introduces a structured iterative framework that integrates principles from software development, assigning roles such as analyst, designer, developer, and tester to guide the generation process. In each iteration, the model refines requirements, adjusts design, and revises code based on feedback from prior outputs. This role-based approach helps the model identify flawed module structures and progressively improve code readability and functional completeness. Addressing the difficulty of model self-diagnosis in the absence of feedback conditions, Self-Debug~\cite{chen2023teaching} introduced ideas analogous to "rubber duck debugging", using few-shot examples to guide the model to perform line-by-line explanation of its own generated code to identify errors, achieving automatic debugging and modification mechanisms without external feedback, significantly improving accuracy and sample utilization efficiency in multiple code tasks. Self-Edit~\cite{zhang2023self} proposed a fault-aware code editor that can perform secondary editing on generated code by LLMs after combining execution feedback information, thereby improving the accuracy of code generation for numerous LLMs. Self-Repair~\cite{olausson2024selfrepairsilverbulletcode} combines code models with feedback models for program repair. The code model generates programs based on user-provided specifications and test cases. If the program fails tests, the feedback model generates explanatory text to help the code model understand error causes and complete repairs accordingly.

To address issues of structural disorganization in complex programs, CodeChain~\cite{le2023codechain} introduces a modular self-revision framework that promotes structured and maintainable code generation. The method guides the model to iteratively refine code by first identifying and clustering representative sub-modules from initial outputs, and then leveraging these sub-modules to enhance subsequent generations. Through this iterative process, the model is encouraged to reuse verified components, thereby improving the modularity, logical coherence, and correctness of the generated programs. Building on this, LeDeX~\cite{jiang2024ledex} enhances the closed-loop self-debugging framework by enabling the model to perform stepwise annotation and analysis of erroneous code, followed by generating repair solutions and verifying their correctness through execution results. Additionally, LeDeX collects data generated during this iterative process to construct high-quality training datasets. These datasets are then used to fine-tune open-source models, significantly improving their code repair capabilities.

Overall, current research has formed a relatively clear technical system in reflection and self-improvement mechanisms: from natural language-level self-feedback to automatic repair combined with execution results, to module-level optimization based on program structure and multi-solution evaluation. These methods not only improve the performance capabilities of single-agent systems in code generation tasks but also provide support for subsequent in-depth research in test-time/inference-time computational scaling and multi-agent collaboration.

\begin{figure*}[h!]
  \centering
  \includegraphics[width=0.99\textwidth]{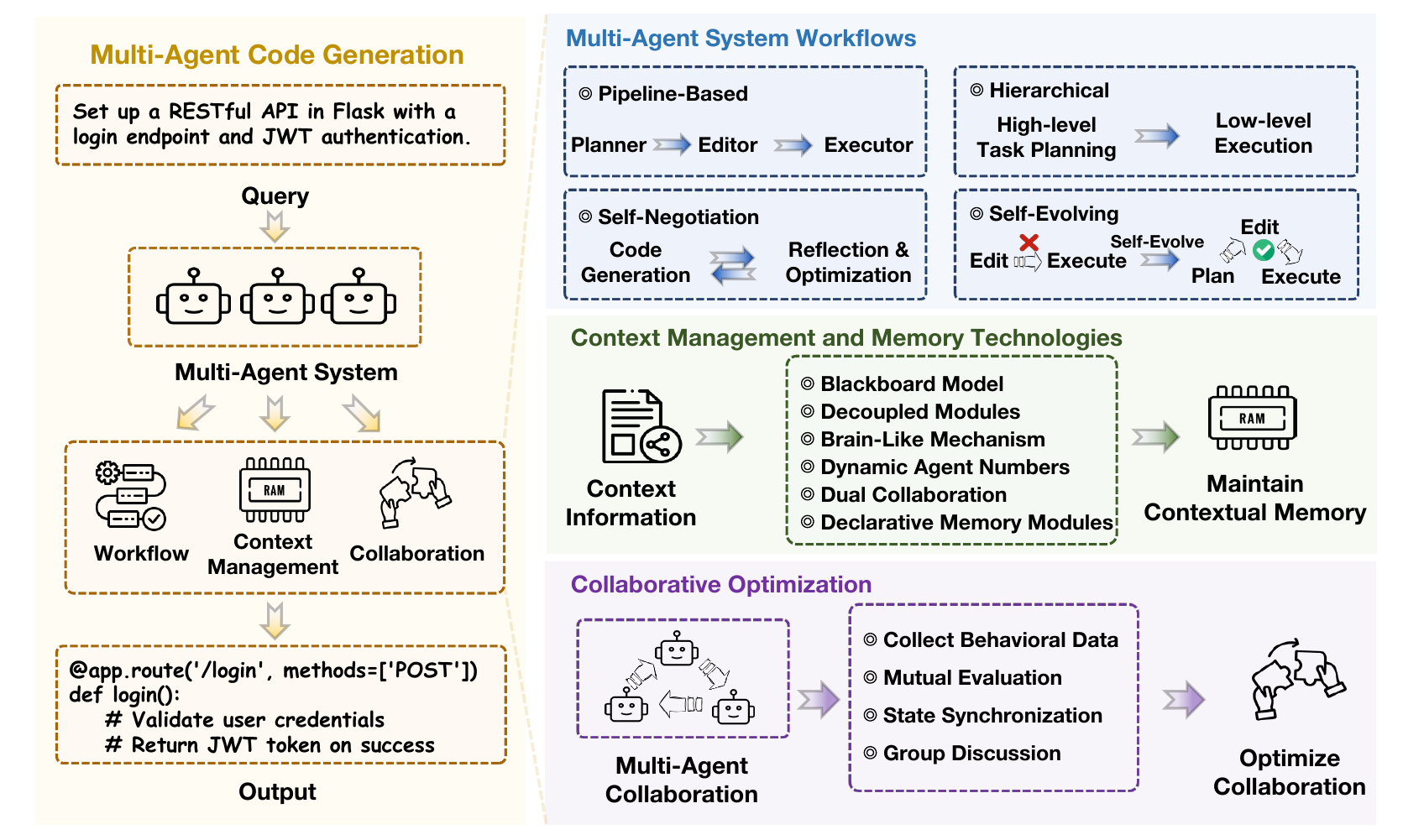}
  \caption{Overview of multi-agent code generation systems}
  \label{fig:multi_agent_overview}
\end{figure*}

\subsection{Multi-Agent Code Generation Systems}

\subsubsection{Multi-Agent System Workflows}

Multi-agent system workflows mainly include four types: pipeline-based labor division, hierarchical planning-execution mechanisms, self-negotiation circular optimization, and self-evolving structural updates.

In the pipeline-based division of labor paradigm, each agent is responsible for a specific stage in the software development process and passes intermediate products to the next agent for processing. The process shows obvious sequentiality. Self-Collaboration~\cite{dong2024self}, which builds upon the traditional waterfall model used in software engineering. It defines a classic three-stage multi-agent process that includes "requirement analysis," "coding implementation," and "testing". In this process, agents work independently at each stage to complete their respective tasks and pass the results to the subsequent stage. Building on this, AgentCoder~\cite{huang2023agentcoder} also constructs a three-stage pipeline system consisting of "programmer", "test designer", and "test executor" agents. This system effectively improves the correctness and executability of code generation through structured task decomposition and test feedback. CodePori~\cite{rasheed2024codeporilargescaleautonomoussoftware} introduces a set of agent roles, including the manager, developer, finalizer, and verifier. In this framework, the manager agent is tasked with parsing natural language requirements and decomposing the tasks. Multiple developer agents work in parallel to write code for different modules, followed by multiple finalizer agents who refine the code. Finally, the verifier agent performs integration testing to ensure the system's functionality.  Additionally, MAGIS~\cite{tao2024magis} tackles repository maintenance tasks and further expands pipeline complexity. It models project managers, maintainers, developers, and quality assurance personnel in the GitHub development process as four agent roles respectively. The system can complete GitHub Issue tracking, assignment, and repair, demonstrating feasibility in real project scenarios. Meanwhile, HyperAgent~\cite{phan2024hyperagent} focuses on cross-language, cross-task code generation problems. This framework transforms natural language requirements into runnable programs through collaboration between planner, navigator, code editor, and executor agents. Meanwhile, it introduces automatic tool chain retrieval mechanisms to improve code generality and transferability. The advantage of this pipeline-based method lies in its clear structure and distinct responsibilities, facilitating system design and debugging. However, they would be limited by serial dependencies when handling complex tasks and lack global feedback.

Unlike sequential pipelines, the hierarchical planning-execution paradigm emphasizes task decomposition by higher-level agents and specific implementation by lower-level agents. PairCoder~\cite{zhang2024pair} simulates collaborative pair programmers, designing two collaborative agents: Navigator for planning and Driver for specific implementation. Navigator is responsible for proposing promising solution methods, selecting the current optimal solutions, and guiding next iterations based on execution feedback. Driver follows the guidance from Navigator for initial code generation, code testing, and refinement. FlowGen~\cite{lin2024soen} simulates various classic software engineering models (such as Waterfall, TDD, Scrum), designing a four-layer agent structure of "requirement engineer", "architect", "developer", and "tester". The system gradually advances development tasks through staged planning and goal verification. Its flexibility under diverse development paradigms makes it a workflow framework with good scalability. Building on this, SoA~\cite{ishibashi2024self} introduces dynamic agent scheduling mechanisms. Based on task complexity and resource usage, the system can spontaneously expand or contract the number of agents, achieving efficient generation and management of large-scale code bases. This self-organizing structure improves system robustness and adaptability, particularly suitable for code construction in heterogeneous task scenarios. Additionally, MAGE~\cite{zhao2024mage} is a multi-layer architecture system for RTL hardware code generation. By decomposing high-level goals into micro-operations and assigning them to different agents, MAGE not only demonstrates the generality of hierarchical structures but also validates the transferability of this paradigm on different types of code (such as Verilog).

In more complex and uncertain tasks, self-negotiation circular optimization mechanisms center on negotiation, reflection, and self-feedback, continuously improving code quality and robustness through multiple rounds of interaction. Some systems choose to have multiple agents evaluate, optimize, and repair candidate solutions in parallel or iterative ways. MapCoder~\cite{islam2024mapcoder} employs a cycle of four specialized agents that work together to recall relevant examples, formulate a solution plan, generate the corresponding code, and then debug any errors. In each iteration, these agents collaborate to produce code and identify defects for repair. By repeating this process multiple times, MapCoder steadily enhances solution quality and maintains stable performance on tasks that require several steps. Subsequently, AutoSafeCoder~\cite{nunez2024autosafecodermultiagentframeworksecuring} designed a framework including coder, static analyzer, and fuzzer, where the coder continuously revises code based on feedback from static security detection by the static analyzer and dynamic security detection by the fuzzer. QualityFlow~\cite{hu2025qualityflowagenticworkflowprogram} organizes a team of LLM agents that first generate unit tests, then apply an LLM-based Quality Checker to imagine whether the code and tests meet the requirements, and finally execute those tests to confirm correctness.  Meanwhile, CodeCoR~\cite{pan2025codecor} introduced a "self-reflection" scoring mechanism. The framework adds reflection agents between code generation, testing, and repair stages to score and locate problems in intermediate products. Then, the results are fed back to preceding agents for the next round optimization. This circular mechanism, based on self-supervised feedback, effectively improved the adaptive capabilities of the model. For high-performance computing fields, MARCO~\cite{rahman2025marco} proposes a multi-agent code optimization framework focusing on improving the performance of LLM-generated code in parallelism, memory efficiency, and architectural adaptability. The system separately sets code generation agents and performance evaluation agents, continuously optimizing code execution efficiency through feedback loops.

As system complexity increases, some work has begun to further explore self-evolutionary mechanisms of multi-agent system structures, allowing systems to spontaneously and dynamically adjust structures and behavioral strategies. SEW~\cite{liu2025sew} proposed workflow self-evolution mechanisms, where systems can dynamically reorganize communication paths and responsibility divisions based on collaboration effectiveness between agents and failure feedback.  This mechanism no longer relies on manually set fixed structures but achieves workflow-level adaptive adjustment through runtime learning and reconstruction. EvoMAC~\cite{hu2024self}, inspired by neural network training processes, compares code generation results with target requirements to obtain text-form feedback from the environment. It designs an innovative "text backpropagation" mechanism to automatically adjust collaboration structures and behavioral strategies between agents.

Additionally, in practical applications, multi-agent systems often improve collaboration effectiveness through "role-playing" mechanisms. Role-playing refers to adding specific identity settings in prompts, such as programmers, testers, project managers, or code reviewers. This mechanism makes agent behavior more consistent with corresponding role responsibilities and thinking patterns. The system designs corresponding prompt strategies for each role, enabling them to perform their duties in collaboration and form clear division of labor processes. For example, agents in the ChatDev~\cite{qian2023chatdev} system respectively play roles of programmers, reviewers, and testers. MetaGPT~\cite{hong2023metagpt} constructs an agent team including roles such as product managers, architects, project managers, and engineers, simulating a complete software company organizational structure.

The implementation of the above workflows and role-playing typically relies on the design of system prompts. The LLM behind each agent understands its current tasks and responsibilities based on preset prompt information. These prompts typically include task division, output requirements, input formats, interface invocation methods, and output formats. This mechanism not only helps language models understand their current task roles but also improves interaction stability by limiting behavioral scope.

\subsubsection{Context Management and Memory Technologies}

Multi-agent systems in code generation tasks frequently rely on sharing and referencing long-range dependency information such as task descriptions, historical modifications, and intermediate products. Therefore, how to effectively maintain a writable, readable, and scalable global context space is an important factor determining system upper limits. Self-Collaboration~\cite{dong2024self} is the first system to introduce the blackboard model into code generation tasks. This method establishes an explicit shared memory space for storing structured information such as task descriptions, intermediate generation results, and code revision records. All agents can read or update blackboard content based on task types as needed and make subsequent decisions accordingly, forming collaboration flows based on shared views. Compared to traditional unidirectional message passing, the blackboard model significantly enhances interaction flexibility between agents.

Subsequently, L2MAC~\cite{holt2023l2mac} draws inspiration from the von Neumann architecture, designing decoupled instruction registers and file storage modules. This system introduces explicit control units to precisely control the scheduling and writing of context content. It organizes context information by program units, effectively breaking through the context window limitations of language models. Therefore, L2MAC enables generation tasks to span longer code structures and demonstrates good context retention capabilities when handling large functions, multi-file projects, and other complex situations. The Cogito~\cite{li2025cogito} system further draws from the three-stage cognition-memory-growth model in neurobiology and designs a brain-like context organization mechanism. Its core idea is to divide context information into three structural categories: short-term memory, long-term knowledge base, and evolutionary growth units. Respectively, these three information categories are responsible for immediate state maintenance during task execution, accumulation of common knowledge, and continuous enhancement of abstract capabilities. This bionic memory system not only enhances the hierarchical sense of context but also provides agents with self-evolution and continuous learning capabilities. Meanwhile, SoA~\cite{ishibashi2024self} introduces a self-organized multi-agent framework that automatically scales the agent pool according to task complexity. Each agent works within its own fixed-size context window, while a central controller keeps subspaces of all agents structurally aligned. By growing the number of agents without increasing individual workloads, SoA prevents information sparsity and confusion, enabling efficient and scalable generation of large codebases.

Besides general technical innovations, some systems have also been designed with more targeted context optimization mechanisms combined with task requirements. GameGPT~\cite{chen2023gamegpt} addresses the problem of frequent repetitive information in multi-round game development tasks. It adopts a "dual collaboration" mechanism to reduce redundant context retransmission, improving context management efficiency. CleanAgent~\cite{qi2024cleanagent} constructed a declarative API memory module based on the Dataprep.Clean library, enabling the system to extract domain knowledge from historical invocation trajectories and reuse it for current tasks, demonstrating the combination of context mechanisms with domain knowledge.

\subsubsection{Collaborative Optimization of Multi-Agents}

As multi-agent systems in code generation tasks have developed increasingly mature, research has begun to focus on how to further improve collaboration capabilities between agents. Such methods can be collectively termed "collaborative optimization". To be specific, they introduce team-level collaboration modeling mechanisms during training to jointly optimize the behavior of multiple agents, thereby improving overall performance metrics. Currently, collaborative optimization remains an emerging research direction with limited related methods and immature technical systems.

An early representative in this direction is Lingma SWE-GPT~\cite{ma2024lingmaswegptopendevelopmentprocesscentric}. Lingma SWE-GPT simulates real software development processes by dividing tasks into three stages: codebase understanding, fault localization, and patch generation, with each stage corresponding to a sub-task agent. To achieve collaborative optimization between agents, Lingma SWE-GPT first collects behavioral data from multiple-stage agents, including reasoning processes, tool invocations, and final results. Then, it optimizes the entire system through supervised fine-tuning, thereby improving multi-stage collaboration quality and overall performance.

CodeCoR~\cite{pan2025codecor} establishes a loop of generation, testing, and repair across four dedicated agents for prompts, code, test cases, and repair suggestions. During training, each agent produces multiple candidates and evaluates outputs from other agents. Then, low-quality prompts, code snippets, tests, and repair advice are pruned based on these mutual assessments. This iterative feedback mechanism sharpens collaboration and yields more reliable code synthesis. Additionally, SyncMind~\cite{guo2025syncmind} takes a system-level view of agent collaboration by tackling the common issue of “out-of-sync” states that arise when multiple agents update a shared codebase. It introduces three complementary evaluation dimensions to study how agents recognize and recover from state drift. These multidimensional mechanisms enrich our understanding of how to optimize collaborative coding workflows.

To alleviate the problem of error propagation during collaboration caused by agent hallucinations,  CANDOR~\cite{xu2025hallucinationconsensusmultiagentllms} coordinates multiple agents, including planners and reviewers. Specifically, CANDOR adopts group discussion strategies to generate accurate reference answers based on consensus among multiple reviewer agents.

\section{Applications of LLM-based Code Generation Agents in Software Development Tasks}

 Recent studies ~\cite{zhou2024language,zhang2023self,schafer2023adaptive,rahman2025marco,jin2024mare} show that code-generation agents have evolved well beyond their traditional role. Today, they support multiple stages of the software-development lifecycle, including automated code writing, debugging and repair, test-code generation, refactoring and optimization, and even requirement clarification.

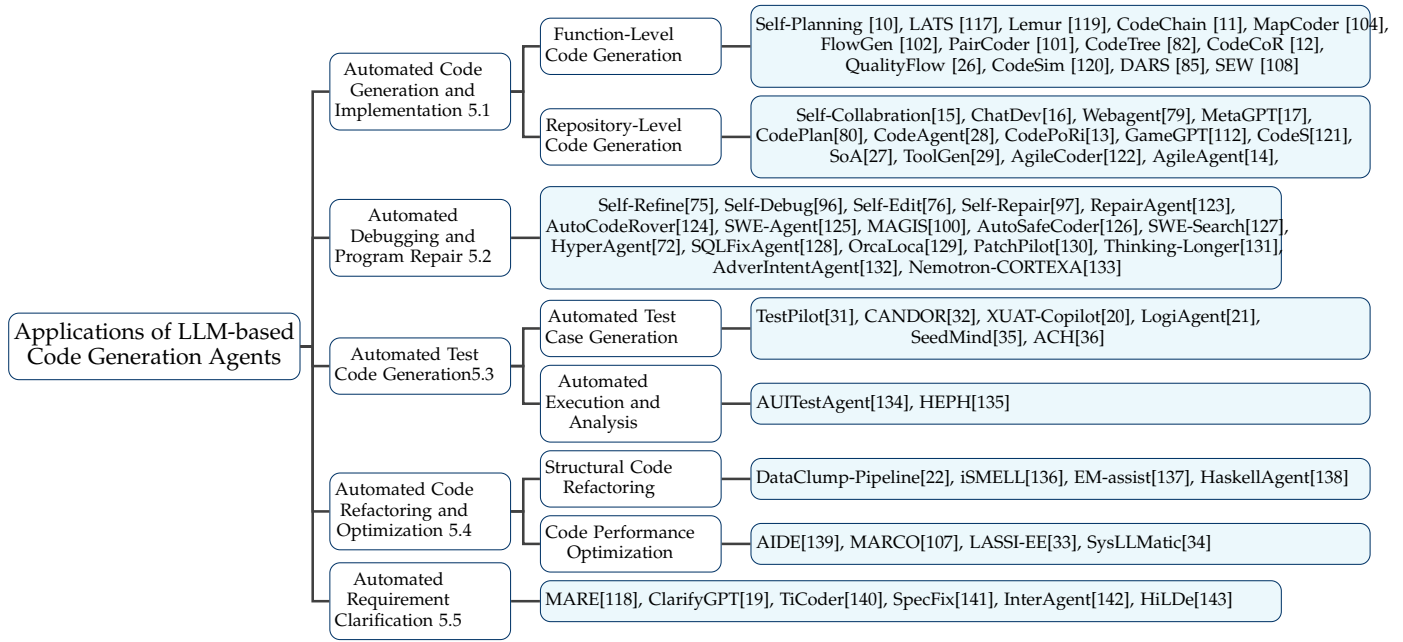
\begin{figure*}[!t]
\centering
    \resizebox{\textwidth}{!}{
        \begin{forest}
            forked edges,
            for tree={
                grow=east,
                reversed=true,
                anchor=center,
                parent anchor=east,
                child anchor=west,
                base=left,
                font=\small,
                rectangle,
                draw=hidden-draw, 
                rounded corners,
                align=center,
                minimum width=4em,
                edge+={darkgray, line width=1pt},
                s sep=3pt,
                inner xsep=2pt,
                inner ysep=3pt,
                ver/.style={rotate=90, child anchor=north, parent anchor=south, anchor=center},
            },
            where level=1{text 
            width=7.0em,font=\scriptsize,}{},
            where level=2{text width=7.0em,font=\scriptsize,}{},
            where level=3{text width=7.0em,font=\scriptsize,}{},
            where level=4{text width=7.0em,font=\scriptsize,}{},
            [
                Applications of LLM-based\\
                Code Generation Agents
                [
                    Automated Code \\ Generation 
                    and \\
                    Implementation \ref{CodeGeneration}
                    [
                        Function-Level \\ Code Generation
                        [
                            Self-Planning \cite{jiang2024self}{,}
                            LATS \cite{zhou2024language}{,}
                            Lemur \cite{xu2023lemur}{,}
                            CodeChain \cite{le2023codechain}{,}
                            MapCoder \cite{islam2024mapcoder}{,}\\
                            FlowGen \cite{lin2024soen}{,}
                            PairCoder \cite{zhang2024pair}{,}
                            CodeTree \cite{li2024codetree}{,}
                            CodeCoR \cite{pan2025codecor}{,}\\
                            QualityFlow \cite{hu2025qualityflow}{,}
                            CodeSim \cite{islam2025codesim}{,}
                            DARS \cite{aggarwal2025dars}{,}
                            SEW \cite{liu2025sew}
                            , leaf, text width=25em
                        ]
                    ]
                    [
                        Repository-Level \\ Code Generation
                        [
                            Self-Collabration\cite{dong2024self}{,}
                            ChatDev\cite{qian2023chatdev}{,}
                            Webagent\cite{gur2023real}{,}
                            MetaGPT\cite{hong2023metagpt}{,}\\
                            CodePlan\cite{bairi2024codeplan}{,}
                            CodeAgent\cite{zhang2024codeagent}{,}
                            CodePoRi\cite{rasheed2024codepori}{,}
                            GameGPT\cite{chen2023gamegpt}{,}
                            CodeS\cite{zan2024codes}{,}\\
                            SoA\cite{ishibashi2024self}{,}
                            ToolGen\cite{wang2024toolgen}{,}
                            AgileCoder\cite{nguyen2025agilecoder}{,}
                            AgileAgent\cite{manish2024autonomous}{,}
                            , leaf, text width=25em
                        ]
                    ]
                ]
                [
                    Automated \\ Debugging 
                    and \\ Program Repair \ref{CodeRepair}
                    [
                        Self-Refine\cite{madaan2023self}{,}
                        Self-Debug\cite{chen2023teaching}{,}
                        Self-Edit\cite{zhang2023self}{,}                        
                        Self-Repair\cite{olausson2024selfrepairsilverbulletcode}{,}
                        RepairAgent\cite{bouzenia2024repairagent}{,}\\
                        AutoCodeRover\cite{zhang2024autocoderover}{,}
                        SWE-Agent\cite{yang2024swe}{,}
                        MAGIS\cite{tao2024magis}{,}
                        AutoSafeCoder\cite{nunez2024autosafecoder}{,}
                        SWE-Search\cite{antoniades2024swe}{,}\\
                        HyperAgent\cite{phan2024hyperagent}{,}
                        SQLFixAgent\cite{cen2025sqlfixagent}{,}
                        OrcaLoca\cite{yu2025orcaloca}{,}
                        PatchPilot\cite{li2025patchpilot}{,}
                        Thinking-Longer\cite{ma2025thinking}{,}\\
                        AdverIntentAgent\cite{ye2025adverintent}{,}
                        Nemotron-CORTEXA\cite{sohrabizadehnemotron}
                        , leaf, text width=30em
                    ]
                ]
                [
                    Automated Test\\
                    Code Generation\ref{CodeTest}
                    [
                        Automated Test \\ Case Generation
                        [
                            TestPilot\cite{schafer2023adaptive}{,}
                            CANDOR\cite{xu2025multi}{,}
                            XUAT-Copilot\cite{wang2024xuat}{,}
                            LogiAgent\cite{zhang2025logiagent}{,}\\
                            SeedMind\cite{shi2024harnessing}{,}
                            ACH\cite{foster2025mutation}
                            , leaf, text width=25em
                        ]
                    ]
                    [
                        Automated \\ Execution  and \\Analysis
                        [
                            AUITestAgent\cite{hu2024auitestagent}{,}
                            HEPH\cite{HEPH}
                            , leaf, text width=25em
                        ]
                    ]
                ]
                [
                    Automated Code \\
            Refactoring and \\ Optimization \ref{CodeRefactor}
                    [
                        Structural Code \\ Refactoring
                        [
                            DataClump-Pipeline\cite{baumgartner2024ai}{,}
                            iSMELL\cite{wu2024ismell}{,}
                            EM-assist\cite{pomian2024assist}{,}
                            HaskellAgent\cite{siddeeq2025distributed}
                            , leaf, text width=25em
                        ]
                    ]
                    [
                        Code Performance \\ Optimization
                        [
                            AIDE\cite{jiang2025aide}{,}
                            MARCO\cite{rahman2025marco}{,}
                            LASSI-EE\cite{dearing2025leveraging}{,}
                            SysLLMatic\cite{peng2025sysllmatic}
                            , leaf, text width=25em
                        ]
                    ]
                ]
                [
                    Automated  \\ Requirement\\
            Clarification \ref{RequirementClarification}
                    [    
                        MARE\cite{jin2024mare}{,}
                        ClarifyGPT\cite{mu2023clarifygpt}{,}
                        TiCoder\cite{fakhoury2024llm}{,}    
                        SpecFix\cite{jia2025automated}{,}
                        InterAgent\cite{vijayvargiya2025interactive}{,}   
                        HiLDe\cite{gonzalez2025hilde}    
                        , leaf, text width=30em
                    ]
                ]
            ]
        \end{forest}}
    \caption{Overview of applications of LLM-based code generation agents in software development tasks}
    \label{fig:Agent_Application_in_SE}
\end{figure*}

\subsection{Automated Code Generation and Implementation}
\label{CodeGeneration}

Automated code generation and implementation have become a key technology for improving software engineering efficiency and quality. With the help of code generation agents, researchers are gradually transforming a large amount of repetitive code writing and debugging work traditionally done manually into automated processes~\cite{zhou2024language,xu2023lemur,liu2025sew,gur2023real}. Currently, the automation level of code generation has expanded from function-level code generation to cross-module, multi-file incremental code evolution, and has further developed into end-to-end project construction oriented toward natural language requirements.

Function-level code generation requires agents to automatically generate functionally correct code segments. Self-Planning~\cite{jiang2024self} is the first work to enable agents to automatically perform task decomposition and formulate execution steps to reduce the coding difficulty of complex requirements. CodeChain~\cite{le2023codechain} achieves modular code generation using chain-based self-revision mechanisms. FlowGen~\cite{lin2024soen} configures a series of different development process models, enabling agents to implement software development modes such as the waterfall model, test-driven development, and agile development according to different role sequences or collaboration processes. PairCoder~\cite{zhang2024pair} adopts a pair programming mode, generating multiple code generation paths through multi-round cycles of agents. CodeTree~\cite{li2024codetree} adopts tree-structured search and verification mechanisms, reducing failure rates when handling complex code generation tasks. CodeCoR~\cite{pan2025codecor} can actively evaluate and reflect on the quality of generated code and initiate automatic correction processes when errors occur. QualityFlow~\cite{hu2025qualityflow} introduces imagination execution mechanisms to achieve rapid quality checking of generated code, saving code execution time in real testing processes. CodeSim~\cite{islam2025codesim} and MapCoder~\cite{islam2024mapcoder} simulate human development, improving correctness and efficiency in long code generation and complex problem-solving processes. DARS~\cite{aggarwal2025dars} can dynamically adjust generation results based on code execution feedback, achieving a higher 47\% Pass@1 rate on the SWE-Bench Lite benchmark compared to previous linear generation or random multi-sampling code generation methods.

Repository-level code generation requires agents to automatically generate and maintain large, complex codebases, including multiple modules and files. One type of work focuses on enabling agents to incrementally add functionality based on understanding existing code structures and module responsibilities. For example, AgileCoder~\cite{nguyen2025agilecoder} and AgileAgent~\cite{manish2024autonomous} progressively develop software systems by simulating human agile development processes. Another type of work hopes that agents can generate code from scratch based on requirements. The Self-Collaboration~\cite{dong2024self} framework enables a single ChatGPT instance to simulate roles at different stages of software development and incorporates software development methodologies, achieving project development without human intervention. ChatDev~\cite{qian2023chatdev} reduces hallucination problems in generated code by having multiple agents communicate through code snippets during the debugging phase. MetaGPT~\cite{hong2023metagpt} simulates real development teams executing multiple development tasks, automatically allocating resources and scheduling development sequences. CodePoRi~\cite{rasheed2024codepori} uses multi-agents to simulate different roles in real software development processes, thereby generating relatively large-scale software systems at low cost. GameGPT~\cite{chen2023gamegpt} achieves automatic construction from natural language descriptions to runnable games, enabling non-programmer users to quickly develop game content through conversational interaction. CodeS~\cite{zan2024codes} overcomes structural confusion and semantic incoherence problems that easily occur in end-to-end large-scale code generation by pre-generating repository structure sketches. SoA~\cite{ishibashi2024self} constructs a multi-code generation agent system that does not require manual central schedulers. Additionally, some work focuses on improving the ability of agents to invoke external tools to better operate codebases. CodeAgent~\cite{zhang2024codeagent} integrates five programming tools, enabling code generation agents to retrieve third-party dependencies in multi-file structures and implement function generation and completion for inter-module calls. ToolGen~\cite{wang2024toolgen} constructs code generation agents with plugin capabilities that can automatically select and invoke corresponding tools based on scenarios and functional needs.

\subsection{Automated Debugging and Program Repair}
\label{CodeRepair}

Automated debugging and repair technology can effectively reduce software defects and improve system stability~\cite{madaan2023self,chen2023teaching,zhang2023self}. The emergence of code generation agents is driving automated program repair from the classic "test-driven patch generation" toward a higher stage of "autonomous defect diagnosis and semantic repair"~\cite{olausson2024selfrepairsilverbulletcode,zhang2024autocoderover,yang2024swe,antoniades2024swe}.

RepairAgent~\cite{bouzenia2024repairagent} provides 14 repair tools commonly used by developers, guiding agent decisions and tool invocations through finite state machines. This system successfully automatically repairs 164 defects on Defects4J. MAGIS~\cite{tao2024magis} uses multi-agents to simulate the planning and coding phases in the GitHub Issue resolution process, improving the GitHub Issue resolution rate to 13.94\% on SWE-Bench. Similarly, HyperAgent~\cite{phan2024hyperagent} uses multi-agents to cover the complete lifecycle from problem decomposition to patch verification, achieving leading performance at the time on multiple benchmarks, including SWE-Bench-Lite, RepoExec, and Defects4J, with good cross-task and cross-language generalization capabilities. SQLFixAgent~\cite{cen2025sqlfixagent} can capture deviations between SQL statements and user intentions, provide diverse repair candidates, and select optimal repair solutions based on failure memories and similar repair histories. AutoSafeCoder~\cite{nunez2024autosafecoder} uses dual feedback from static security detection and fuzzing dynamic security detection to guide agents in continuously revising code, reducing vulnerability rates by approximately 13\% on SecurityEval and improving code security. OrcaLoca~\cite{yu2025orcaloca} improves problem localization accuracy and overall repair effectiveness in complex code repositories by introducing strategies such as priority scheduling, action decomposition, and context pruning. PatchPilot~\cite{li2025patchpilot} can use efficient localization strategies to quickly find vulnerable code lines and generate verifiable repair patches with fewer invocations and token consumption, reducing the cost of program repair. Thinking-Longer~\cite{ma2025thinking} enables small-scale code generation agents to approach the vulnerability repair capabilities of larger-scale agents by increasing thinking complexity, achieving low-resource deployment of automated program repair systems. AdverIntentAgent~\cite{ye2025adverintent} avoids overfitting problems in generated patches by constructing adversarial test cases for each possible program intention. Nemotron-CORTEXA~\cite{sohrabizadehnemotron} adopts diverse repair candidate mechanisms for Python program repair.

\subsection{Automated Test Code Generation}
\label{CodeTest}

With the popularization of code generation technology, the focus of software development is shifting toward verification and testing. Moreover, the testing phase has evolved into a key aspect for managing and verifying large amounts of generated code. LLM-based code generation agents can generate unit tests, integration tests, and security test cases based on requirements, code, and related documentation, as well as execute automated testing processes~\cite{schafer2023adaptive,hu2024auitestagent,Liu2024TestCase,zhang2025logiagent,hu2024auitestagent}.

In unit testing, traditional search methods like EvoSuite \cite{EvoSuite} typically rely on structural information and code execution paths, unable to understand function semantics, and can only attempt to collide with boundaries through extensive coverage or heuristic exploration. In contrast, LLMs can better understand functional intentions and semantic boundaries. TestPilot~\cite{schafer2023adaptive} can automatically generate test cases for JavaScript APIs, achieving 52.8\% branch coverage on 25 npm packages and 1,684 API functions, improving 27\% compared to the feedback-driven tool Nessie. CANDOR~\cite{xu2025multi} can generate Java unit test cases end-to-end, successfully surpassing EvoSuite in metrics such as line coverage. In integration testing, XUAT-Copilot~\cite{wang2024xuat} can automatically complete the determination of test instruction types, specific parameter filling, and effect verification. The system achieves automated integration testing for mobile payment applications such as WeChat Pay. LogiAgent~\cite{zhang2025logiagent} integrates components for test scenario generation, API request execution and verification, and logic notification. It overcomes the problem that traditional REST API testing only focuses on detecting server crashes and error codes. Meanwhile, this system achieves automated detection of a series of logic problems brought by business development and domain-specific requirements. In security testing, LLM-based code generation agents can automatically derive aggressive input patterns and boundary condition cases from natural language documentation, interface definitions, and code structures. SeedMind~\cite{shi2024harnessing} automatically generates high-quality seed inputs through LLMs as starting points for gray-box fuzzing. The framework can dynamically adjust generation strategies based on test feedback to improve test coverage. Meta's ACH~\cite{foster2025mutation} system generates mutants that conform to program semantics through LLMs and judges mutant equivalence, improving the efficiency of mutation testing.

Automated testing processes involve not only test case generation but also execution, evaluation, and analysis. For example, AUITestAgent~\cite{hu2024auitestagent} can automatically identify interactive instructions and execute full-process graphical user interface operations and functional verification through pure natural language testing requirements. The HEPH system \cite{HEPH} from NVIDIA achieves verification of correctness of C/C++-based embedded systems in various scenarios by analyzing requirement documents and retrieving various related files, and guides the next round of test case generation with coverage reports after each testing round.

Although the automated testing capabilities of LLM-based code generation agents have not yet comprehensively surpassed traditional tools, their semantic understanding advantages in covering boundaries and discovering hidden errors deserve future in-depth exploration.

\subsection{Automated Code Refactoring and Optimization}
\label{CodeRefactor}

Automated code refactoring and optimization help improve code maintainability and runtime efficiency. LLM-based agents can understand the semantics of existing code and combine external tools such as static analysis, test feedback, and performance monitoring to decide how to rewrite code~\cite{jiang2025aide,rahman2025marco}.

In structural code refactoring, Baumgartner et al.~\cite{baumgartner2024ai} propose a modular automated refactoring pipeline based on LLMs for detecting and repairing data clumps in GitHub repositories. EM-Assist~\cite{pomian2024assist} designs plugins for overly long functions written by users in Java and Kotlin, using LLMs to provide extract method suggestions. iSMELL~\cite{wu2024ismell} can automatically invoke multiple professional detection tools for different types of code smell and integrate tool detection results into LLMs to generate targeted refactoring suggestions. Siddeeq et al.~\cite{siddeeq2025distributed} propose a multi-agent system for automated Haskell code refactoring, completing tasks such as variable renaming, function extraction, redundancy elimination, module reorganization, and style unification. The system solves the code refactoring difficulties brought by Haskell's pure functional characteristics.

In code performance optimization, both LASSI-EE~\cite{dearing2025leveraging} and SysLLMatic~\cite{peng2025sysllmatic} use LLMs to analyze various performance and energy consumption metrics during code runtime, achieving improvements in software system performance and energy consumption through multi-round diagnosis and optimization.

\subsection{Automated Requirement Clarification}
\label{RequirementClarification}

Requirement clarification aims to eliminate ambiguity and uncertainty in instructions to ensure accurate understanding of the true intentions from the users. LLM-based code generation agents can handle ambiguous or incomplete natural language requirements and gradually clarify them under user feedback guidance, thereby generating code that meets final requirements~\cite{jin2024mare,fakhoury2024llm}.

Existing applications have evolved along a developmental trajectory that progresses from static understanding of requirements to dynamic dialogue-based clarification. This path includes the detection of ambiguity, the proactive generation of clarification questions, and ultimately, the collaborative control of generation processes between humans and machines. ClarifyGPT~\cite{mu2023clarifygpt} judges whether ambiguous requirements exist by checking the consistency of multiple generation results from LLMs and obtains accurate requirements through question-and-answer with users. Furthermore, combining test-driven development concepts, TiCoder~\cite{fakhoury2024llm} guides users to clarify natural language requirements by automatically generating test cases, thereby improving code generation accuracy. SpecFix~\cite{jia2025automated} can automatically guide LLMs to make minimal modifications to original requirement texts to eliminate unnecessary ambiguity. However, due to lack of interaction with the outside world, overall effectiveness is limited by LLM capabilities. InterAgent~\cite{vijayvargiya2025interactive} verifies the capabilities of LLM-based code generation agents in identifying unclear instructions, actively clarifying, and improving performance through interaction in complex tasks. The framework ultimately finds that under incomplete requirement conditions, most LLMs perform poorly without active interaction but can effectively obtain key information when interactive, approaching performance under complete input conditions. Additionally, HiLDe~\cite{gonzalez2025hilde} enables human users to integrate personalized requirements into key decision positions of code generation by selecting subsequent code generation results.

\section{Evaluation Methods and Benchmarks}

Evaluating code generation agents, particularly those used to solve complex software engineering tasks, is a fundamental and extremely challenging problem. The core challenge lies in the fact that evaluation methods must go beyond judging code syntax or pass rates to deeply explore the problem-solving capabilities of agents in complex, dynamic software development scenarios. This chapter provides a structured overview of existing evaluation methods and benchmarks for code generation agents. These methods have evolved significantly over time. Initially, they focused on assessing self-contained code snippets, such as individual functions. Gradually, however, the focus has shifted toward evaluating the ability of agents to perform multi-step, interactive tasks within large and complex software repositories. This shift reflects the expanding boundaries of code generation capabilities in the field of software engineering.

\subsection{Evaluation Benchmarks}

The evaluation of code generation tasks heavily relies on standardized benchmark datasets. Early benchmarks mainly focus on basic independent unit code generation tasks, while newly developed benchmarks in recent years are committed to simulating real software development tasks, emphasizing the interaction, planning, and execution capabilities of agents in complex environments. Existing evaluation benchmarks mainly include three major categories: method/class-level, programming contest-level, and benchmarks oriented toward real software development scenarios.

\textbf{Method/Class-level Code Generation Benchmarks}: In the initial phase of research on code generation agents, evaluation primarily focused on independent and clearly specified programming tasks. The central aim was to determine whether models could transform natural language descriptions into syntactically and functionally correct code snippets. This line of work is typically referred to as method-level or class-level code generation, in which models receive high-level functional requirements, optionally accompanied by method signatures, and are required to generate complete method bodies.
The benchmarks constructed during this period provide methodological foundations for later research targeting more complex tasks. They also establish an evaluation framework based on functional correctness, which has been widely adopted in both academic and practical settings.

Among numerous benchmarks, \textbf{HumanEval}~\cite{chen2021evaluating}, launched by OpenAI in 2021, has milestone significance. It borrows ideas from online programming platforms (such as LeetCode), providing unit test cases for each problem. This benchmark shifts evaluation standards from vague text similarity to objective, automatable code execution testing. It establishes a new gold standard for the code generation field. Similarly, \textbf{MBPP}~\cite{mbpp} dataset proposed by Google Research focuses on entry-level Python programming tasks. In this benchmark, each problem includes natural language descriptions, reference answers, and three assertion statements for automatic testing, covering various problem types from mathematical calculations to basic data processing.

\textbf{Programming Contest Code Generation Benchmarks}: As the capabilities of LLMs improve, code generation evaluation gradually shifts from simple function implementation to complex tasks requiring deeper logical reasoning and algorithmic design, such as programming contest problems. Such tasks not only require models to master programming language syntax but also test their abilities to analyze problems, design algorithms, and organize complex logic. The academic community has constructed several high-quality programming contest benchmarks. Among them, \textbf{APPS}~\cite{apps}, \textbf{CodeContests}~\cite{li2022competition}, and \textbf{LiveCodeBench}~\cite{livecodebench} are  three most representative benchmarks. APPS collects 10,000 problems from online evaluation websites such as Codeforces and divides them into three difficulty levels. CodeContests contains 13,328 problems collected from real programming contests, covering C++, Python, and Java languages. To avoid data contamination risks \cite{DataContamination}, LiveCodeBench continuously maintains new programming contest problems from LeetCode, AtCoder, and CodeForces platforms, updating the dataset every few months. As of now, it includes 1,055 problems from May 2023 to May 2025. Evaluation of these benchmarks typically uses test case pass rates as core metrics.

\textbf{Benchmarks for Real Software Development Scenarios}: With the rapid development of code generation agents, research focus has shifted toward building autonomous agents capable of executing complex tasks in real software projects. A series of benchmarks aimed at simulating real software engineering scenarios has emerged. A key characteristic of these benchmarks is that they provide a complete software development environment. In this environment, agents are required to solve real-world tasks by interacting with codebases, command-line interfaces, debugging tools, and other resources, simulating the workflow of human developers. For example, \textbf{SWE-Bench}~\cite{swebench} is a large-scale GitHub Issues resolution benchmark containing thousands of real GitHub Issues. The Issues are collected from 12 popular Python codebases, with task types covering defect repair and new feature development. Due to the complexity of the original dataset, researchers have also constructed two important derivative versions. One is \textbf{SWE-Bench-Lite}, which screens 300 more self-contained functional defect repair tasks for faster, more focused evaluation. The other is \textbf{SWE-Bench-Verified}, where OpenAI researchers manually verify and screen instances from the original SWE-Bench, ultimately forming a test set containing 500 high-quality instances. \textbf{CodeAgentBench}~\cite{zhang2024codeagent} contains 5 real Python software projects and 101 tasks, requiring models to implement new functions based on functional documentation and existing project code. \textbf{Web-Bench}~\cite{webbench} covers multiple web project development tasks from scratch. This sequential, project-based approach differs from other benchmarks like CodeAgentBench and SWE-Bench, focusing on evaluating the long-term planning, memory, and sustained contextual understanding capabilities of agents across multiple steps in project development tasks. \textbf{Aider}~\cite{aider} contains 225 high-difficulty educational programming exercises covering multiple programming languages, requiring models or agents to implement corresponding functions based on instruction documentation. Additionally, there are other real project-oriented code generation benchmarks, such as \textbf{EvoCodeBench}~\cite{evocodebench} and \textbf{DevEval}~\cite{deveval}, which require models to generate target code that meets expectations given complete project contexts and requirements.

\subsection{Evaluation Metrics}

Code generation agent evaluation mainly includes the following aspects: functional correctness, process efficiency, and non-functional software quality.

\textbf{Functional Correctness Metrics}: Functional correctness metrics are the most core dimension for evaluating code generation capabilities \nocite{CodeScore}. They are typically measured by executing generated code and testing whether its output meets expectations. Pass@k~\cite{chen2021evaluating} is the most commonly used metric for calculating functional correctness. For each problem, the model independently generates $k$ candidate code samples, and if at least one sample can pass all unit tests, the problem is considered successfully solved. Pass@k measures the probability of success at least once in $k$ attempts. To eliminate sampling bias, the following unbiased estimator is typically used for calculation:
\begin{equation*}
\text{pass@k} = \mathbb{E}_{\text{problems}} \left[ 1 - \frac{\binom{n-c}{k}}{\binom{n}{k}} \right]
\end{equation*}
where $n$ is the total number of samples generated for a single problem ($n \ge k$), and $c$ is the number of correct samples passing tests. This metric simulates scenarios where developers might generate multiple alternative solutions and select one in actual work. Therefore, better reflecting the practical value of models compared to Pass@1, which only evaluates single generation.

\textbf{Evaluation Metrics for Code Generation Agents}: 
In recent years, researchers have proposed specialized evaluation metrics for code generation agents. These metrics aim to evaluate agents from multiple dimensions, including output quality, efficiency, cost, and execution process. Among them, task success rate~\cite{swebench} is a fundamental metric. It directly measures whether an agent can solve a given problem. For example, in SWE-bench, task success is defined as whether the agent-generated patch passes all relevant test cases.

However, task success rate alone cannot fully reflect the real-world performance of the agent. Efficiency and cost must also be taken into account. This includes the API call cost and token consumption~\cite{hu2025optimizingtokenconsumptionllms} required to complete a task, which are directly related to economic cost. It also includes latency from receiving a task to producing a solution, which affects user experience. In addition, local model deployment may incur computational resource costs~\cite{zhao2025insightsdeepseekv3scalingchallenges}.

Besides final outputs, the execution trajectory of the agent is also an important evaluation target. Trajectory efficiency reflects how many steps the agent takes to complete the task. An efficient agent should reach the goal through a minimal and effective sequence of actions, avoiding redundant operations. Some metrics also evaluate the quality of intermediate steps, such as tool usage accuracy~\cite{swebench}. This refers to whether the agent can select the correct tools and use them with valid parameters.

\textbf{Other Software Quality Evaluation Metrics}: As code generation agents become more capable, evaluation criteria have gradually expanded to include non-functional attributes that are essential to software quality, such as security. Emerging benchmarks, such as SEC-bench~\cite{lee2025secbenchautomatedbenchmarkingllm}, are designed to assess the ability of agents to repair real-world security vulnerabilities.
Comprehensive evaluation also considers code quality~\cite{sealign}, which involves several aspects. Static analysis tools can be used to measure code readability and complexity, for example, by calculating cyclomatic complexity and lines of code~\cite{siddeeq2025llmbasedmultiagentintelligentrefactoring}. Maintainability can be assessed using metrics such as inter-module coupling~\cite{siddeeq2025llmbasedmultiagentintelligentrefactoring}. In addition, agents that follow good engineering practices are expected to update related test cases during development to maintain high test coverage~\cite{chen2024evaluatingsoftwaredevelopmentagents}.

\section{Deployed Code Generation Agent Tools}

Several multi-agent code generation tools have been deployed and have produced widespread impact in the current market. We roughly categorize them into three types to show their evolutionary trajectory: 1) \textbf{Co-pilot}: Human-machine close collaboration, serving as the co-pilot of developers to assist programmers in coding. 2) \textbf{Collaborator}: Capable of understanding entire codebase contexts and engaging in deep interactive collaboration with developers. 3) \textbf{Autonomous Team}: Aimed at automating the entire development process, where humans play more of a client or manager role. Currently, emerging code generation agent tools in the market embody this evolutionary trajectory, playing roles at different stages of programming assistant, collaborative partner, and autonomous development team. Table \ref{tab:agent_comparison} presents a horizontal comparison of these mainstream tools.

\begin{table*}[t]
\centering
\scriptsize 
\setlength{\tabcolsep}{5pt} 
\renewcommand{\arraystretch}{1.3} 
\caption{Horizontal Comparison of Code Generation Agent Tools}
\label{tab:agent_comparison}
\begin{tabularx}{\textwidth}{lXXXXXX}
\toprule
\textbf{Tool Name} & \textbf{Core Paradigm} & \textbf{Primary Integration Method} & \textbf{Context Engine} & \textbf{Best Application Scenario} & \textbf{Main Advantages} & \textbf{Known Limitations} \\
\midrule
\textbf{GitHub Copilot} & Programming Assistant & IDE Extension Plugin & Retrieval Augmented & Code Completion & Wide adoption, excels at routine tasks & Requires human supervision, may overlook global context \\
\textbf{Devin} & Autonomous Development Team & Independent Online Platform & Sandbox-based Knowledge Base & End-to-End Code Generation & High autonomy, handles full task lifecycle & Low practical reliability, unpredictable results, prone to loops \\
\textbf{Cursor} & Collaborative Partner & AI-Native IDE & Vector Semantic Indexing & Code Completion, Code Development, Codebase Q\&A & Deep codebase understanding, excellent interaction experience & Weaker autonomous agent capabilities \\
\textbf{Tongyi Lingma} & Collaborative Partner & Enterprise Platform Integration & Repository Knowledge Graph & Code Development & Robust and systematic exploration of solution space & More focused on enterprise-level applications \\
\textbf{Claude Code} & Autonomous Development Team & Command Line Interface & Ultra-Long Context Window + Agent Search & End-to-End Code Generation, Codebase Q\&A & Powerful code agent, global codebase understanding & Potentially high cost \\
\bottomrule
\end{tabularx}
\end{table*}

\textbf{GitHub Copilot}~\cite{copilotgithub}: As an integrated programming agent, GitHub Copilot has elevated programming assistance to new heights. Its technical core lies in using Retrieval-Augmented Generation (RAG) to dynamically construct contexts and safely execute tasks in cloud-based sandboxes powered by GitHub Actions. This architecture enables deep integration into the GitHub ecosystem (such as Issues, PRs), not only assisting in coding but also automating workflows from requirements to code commits.

\textbf{Devin}~\cite{devin}: Devin was first released in early 2024, promoted as the first AI software engineer. It aims to fully automate various complex software engineering tasks by empowering LLM-based agents with the ability to use terminal tools, code editors, and web browsers, covering the entire workflow from planning, execution, to debugging and deployment. However, actual use revealed numerous problems with Devin, including low success rates, frequent loops, and difficult-to-resolve hallucination issues. This indicates that current code generation agents still face enormous difficulties when handling various complex edge cases in real-world software engineering projects.

\textbf{Cursor}~\cite{cursor}: Cursor is an AI-native standalone IDE and a deep practice of the collaborative partner concept. Through deep customization of VS Code, it achieves a local-first agent architecture. Its key technology is persistent vector embedding indexing of codebases, achieving low-latency global context awareness. This enables its agents to directly interact with local file systems and terminals, demonstrating efficient, deep interaction and collaboration capabilities with developers when handling complex code refactoring tasks spanning multiple files.

\textbf{Tongyi Lingma}~\cite{lingma}: Tongyi Lingma is a collaborative partner based on  CodeQwen, with its LingmaAgent framework showing a unique technical approach. It constructs codebase knowledge graphs and innovatively uses Monte Carlo Tree Search (MCTS) for systematic navigation and fault localization, achieving deep understanding of complex software repositories. This methodology enables it to perform well in advanced collaborative tasks such as multi-file dependency analysis and automated program repair, rather than just simple code completion.

\textbf{Claude Code}~\cite{claudecode}: Claude Code is a terminal-native programming agent whose design significantly advances toward the autonomous development team direction. Its technical core is using ultra-long context windows of up to 200K tokens for holistic semantic understanding of entire codebases. Combined with its hybrid reasoning engine and extended thinking mode, it can autonomously plan and execute complex code generation, refactoring, and debugging tasks starting from highly abstract natural language requirements. Therefore, Claude Code is pushing human roles more toward requirement definition and final supervision.

\section{Challenges and Future Directions}

We discuss and summarize the challenges currently faced by LLM-based code generation agents and future development directions from five dimensions.

\subsection{Limitations of Agent Core Capabilities}

\textbf{Handling Domain-Specific Tasks}: Current LLM-based agents exhibit significant limitations when dealing with tasks requiring deep domain knowledge and complex professional reasoning. This problem is particularly prominent when tasks involve domain-specific terminology or industry standards. Due to the lack of structured domain knowledge bases and specialized training, agents can easily produce domain-related understanding biases. Moreover, they may also experience logical collapse or even generate hallucinations when handling problems beyond their knowledge boundaries. To break through this bottleneck, effective domain knowledge enhancement methods need to be explored.

\textbf{Intent Understanding and Context Awareness}: Human instructions often have informal and ambiguous characteristics. When task objectives themselves are vaguely defined or depend on implicit contexts, agents without strong intent understanding capabilities are prone to misunderstandings, leading to outputs that do not meet expectations. Therefore, code agent systems need to improve their intent reasoning and context awareness capabilities. Meanwhile, they need to introduce interactive clarification mechanisms to ensure alignment with human goals.

\textbf{Handling Large-Scale, Complex Codebases and Project-Level Dependencies}: When processing real projects containing massive files and complex dependency relationships, existing code generation agents have limited capabilities in long-context modeling, cross-file dependency analysis, and overall software architecture understanding, greatly limiting their application effectiveness in large projects.

\textbf{Multimodal Understanding and Generation}: Modern software development is a multimodal process, often involving non-textual information such as UI design sketches, architecture diagrams, and flowcharts. Current LLM-based agents cannot fully understand and utilize multimodal information for code generation, which is particularly critical in frontend development and user interface implementation tasks. This is one of the important research directions for the future.

\textbf{Context Engineering}: In real-world software development scenarios, information is often scattered across multiple files, modules, and documents. Without a systematic mechanism for constructing and managing context, models are highly prone to generating inconsistent or even incorrect code. This issue becomes particularly prominent in multi-step tasks. Empirical evidence suggests that many failures of intelligent agents do not stem from the models themselves, but rather from defects in the ingested context. These defects can be categorized into four typical problems: context poisoning (incorrect or hallucinated information contaminates subsequent reasoning), context distraction (redundant information overwhelms key signals), context confusion (format inconsistencies lead to misinterpretation), and context conflict (contradictory information causes decision-making errors). Although existing studies have attempted to mitigate this challenge through approaches such as retrieval augmentation and long-term memory modules, there is still a lack of stable and practical solutions for software development. Therefore, context engineering is not merely an extension of prompt design, but rather a dynamic methodology whose core objective is to deliver the right information and tools to the model—at the right time and in the appropriate format. Only through systematic organization and optimization of context can intelligent agents maintain consistency, reliability, and efficiency when operating in complex tasks and large-scale systems.

\subsection{Robustness and Updatability Challenges of Agent Systems}

\textbf{Error Cascading Within Agent Systems}: In multi-agent systems, minor deviations from upstream agents (e.g., an incorrect API call parameter) are used as inputs for downstream agents. If subsequent agents fail to identify and correct this error, they will make further decisions based on the erroneous information. Such deviations are amplified level by level along the collaboration chain, forming error cascading effects that may ultimately lead to systemic failure of entire tasks. This error cascading effect is an important factor threatening the reliability of multi-agent systems.

\textbf{Coordination and Management Complexity in Multi-Agent Collaboration}: As the number of agents in a system increases, potential interaction relationships between them grow exponentially, causing system complexity to rise sharply. Without efficient coordination mechanisms, collaboration among numerous autonomous agents can easily fall into chaos, causing problems like communication bottlenecks, responsibility ambiguity and goal drift. Ultimately, the chaotic collaboration will damage overall system stability and task execution efficiency. How to design a robust collaboration framework to effectively manage large-scale agent teams is a complex systems engineering challenge.

\textbf{Agent Knowledge Updates and Continuous Learning}: Software development environments are inherently dynamic, with new programming language features and framework versions constantly emerging. However, code-generation agents trained in a one-off manner possess fixed knowledge bases, making them unable to keep up with the latest technologies and inevitably prone to obsolescence. Thus, agents need the capability to actively and continually acquire new knowledge. Efficient fine-tuning methods such as LoRA \cite{hu2021loralowrankadaptationlarge} can incrementally adjust model parameters, but their effectiveness in learning new knowledge remains limited and they inevitably suffer from degradation of previously acquired knowledge. Retrieval-augmented generation (RAG), on the other hand, relies on external vector databases for real-time retrieval, yet it requires continuous maintenance of external knowledge stores and lacks deep integration between retrieved content and the internal knowledge of the model. Therefore, an effective continual learning mechanism that can efficiently incorporate new knowledge into agents while avoiding catastrophic forgetting is still missing.

\subsection{Challenges of Tool Integration and Deployment for Agents in Open Environments}

\textbf{Flexibility and Security in Tool Usage}: To accomplish meaningful tasks, agents must be equipped with the ability to use external tools (e.g., compilers, testing frameworks, database interfaces). However, providing agents with a predefined and static toolset lacks flexibility and cannot meet the needs of open and dynamic real-world tasks. Thus, an important research question is how to enable agents to move beyond the constraints of static tools, achieving on-demand and flexible tool discovery and integration, while simultaneously ensuring the security of tool execution to prevent potential malicious operations.

\textbf{High Operational Costs of Agent Systems}: Multi-agent systems enhance problem-solving capabilities through iterative collaboration and interaction, but this often comes at the expense of significant computational and time costs. Each interaction among agents and every call to a large language model incurs substantial overhead. As task complexity increases and interaction rounds grow, the overall cost escalates sharply, becoming a major economic and technical barrier to large-scale deployment of multi-agent systems in real-world development scenarios.

\textbf{Lightweight Deployment on Edge Devices}: Deploying code-generation agents on edge devices such as smartphones and drones can reduce reliance on cloud servers, better satisfying  needs of users for real-time response and autonomy. However, the limited computational and storage capacities of such devices pose challenges for lightweight deployment of agent systems. Future solutions will require model compression and inference acceleration techniques to reduce computational and storage overhead while preserving generation capabilities. Examples include quantization to lower representation costs of model weights, knowledge distillation to transfer capabilities from large models to smaller ones, and inference pruning to shorten inference length.

\subsection{Trustworthiness, Security, and Ethical Risks of Agent Systems}

\textbf{Reliability and Debuggability of Agent Systems}: The inherent hallucination problem of large language models makes them prone to generating content that appears plausible but is in fact false, thereby undermining the reliability of agent systems. If agent systems cannot consistently and reliably provide accurate information, they become extremely dangerous in high-risk scenarios (e.g., critical infrastructure control). Furthermore, when agent behavior deviates from expectations, their complex internal decision-making chains and nondeterministic outputs make debugging and troubleshooting particularly difficult.

\textbf{Malicious Code Generation}: As the capabilities of agent systems improve, their security risks require greater attention. On the one hand, users may craft prompts to bypass model safety mechanisms and induce the generation of malicious code. On the other hand, agents may memorize and reproduce historical data containing security vulnerabilities or attack scripts, thereby automatically generating malicious code targeting specific systems. To mitigate such risks, comprehensive security review mechanisms must be introduced before and after the generation process \cite{hua2024trustagent}. Potential measures include conducting interpretability analyses of generation paths to identify risks prior to code output, and applying sandbox-based detection to sensitive operations in the generated content afterward.

\textbf{Code Copyright and Ownership}: Code-generation agents are trained on public code data from open-source repositories, and may also involve private code data from enterprises and individuals. Consequently, agents may generate code segments that closely resemble copyright-protected code, or reproduce open-source content subject to licensing restrictions that is then used for commercial purposes—leading to copyright infringement risks. Moreover, due to the current lack of unified standards for copyright recognition of agent-generated content, the legal responsibilities associated with using such systems in software development remain ambiguous and prone to disputes. Future research needs to incorporate techniques such as “digital watermarking” \cite{zhang2025robust} during both training and inference to trace the origin of generated content, alongside establishing robust community-level ethical guidelines.

\subsection{Evaluation System Completeness and Software Paradigm Evolution}

\textbf{Inadequate Effectiveness and Comprehensiveness of Evaluation Methods}: Existing evaluation methods for code generation systems have clear limitations. They often focus solely on metrics like test pass rates (e.g., Pass@k), while overlooking important factors such as human cognitive load and the effort required for intervention in real collaborative scenarios. This one-dimensional approach fails to capture the practical utility of such systems. There is a pressing need for a more comprehensive evaluation framework that considers multiple aspects, including: (1) task effectiveness and efficiency, (2) the quality of human-agent interaction, (3) system security and interpretability, and (4) user experience and cognitive demands. Such a framework is essential to support the development of code generation agents that are not only effective but also trustworthy and user-friendly in real-world applications.

\textbf{Software Paradigm Transformation}: The current software development paradigm is characterized by a collaborative model in which human developers work alongside code generation agents. In this model, agents function as advanced productivity tools embedded within the traditional software engineering workflow. Human developers remain the central decision-makers, leveraging agents to assist with tasks such as coding, debugging, and maintenance. The end product continues to be conventional software delivered as static artifacts. However, a shift toward a new paradigm is emerging, where agents no longer serve merely as supporting tools but take on a more autonomous role in delivering complete task outcomes. In this future paradigm, users interact with the system through high-level intent descriptions rather than low-level code manipulation, and their role evolves from that of software constructors to providers of problem statements. The agent system interprets these intents, dynamically generates the necessary code, executes actions internally, and directly returns results, effectively offering software as a service on demand. This transformation imposes new and significant demands on the  autonomy, reliability, and ability of agents to complete complex tasks in an end-to-end manner.

\section{Conclusion}

This paper investigates code generation with LLM-based agents as a new paradigm in software development. It highlights three key features that set them apart from traditional code generation methods: autonomy, broader task coverage, and engineering practicality. We review the development path of this technology and analyze its core components from a methodological perspective. At the application level, we summarize typical use cases across the software development lifecycle and list representative tools that have been deployed. We also discuss current technical challenges, including integration with development environments and ensuring code quality, which remain central problems for future research.

Looking ahead, we are confident that with the gradual resolution of existing challenges, code generation agents will fundamentally transform the field of software engineering. These agents promise to free developers from repetitive and tedious coding tasks, enabling them to focus more on creative problem formulation and system-level design. We sincerely hope that this survey serves as a valuable resource for peers in the community, sparks further innovative thinking, and contributes to the advancement and maturity of this promising technology.

\ifCLASSOPTIONcompsoc
  \section*{Acknowledgments}
\else
  \section*{Acknowledgment}
\fi

We would like to thank Tianchi Xu and Mengren Zheng for their contributions to our early work.

\ifCLASSOPTIONcaptionsoff
  \newpage
\fi

\bibliographystyle{IEEEtran}
\bibliography{reference_english}

\end{document}